\documentclass[iop]{emulateapj}
\newcommand{\lya}{Ly$\alpha$}
\newcommand{\ha}{H$\alpha$}
\newcommand{\hb}{H$\beta$}
\newcommand{\oiii}{[O\,{\sc iii}]}

\begin{document}

\title{PHYSICAL PROPERTIES OF SPECTROSCOPICALLY-CONFIRMED GALAXIES AT $z\ge6$. 
III. STELLAR POPULATIONS FROM SED MODELING WITH SECURE \lya\ EMISSION AND
REDSHIFTS\footnotemark[$\ast$]}
\footnotetext[$\ast$]{
Based in part on observations made with the NASA/ESA Hubble Space Telescope,
obtained from the data archive at the Space Telescope Science Institute, which
is operated by the Association of Universities for Research in Astronomy, Inc.
under NASA contract NAS 5-26555.
Based in part on observations made with the Spitzer Space Telescope, which is
operated by the Jet Propulsion Laboratory, California Institute of Technology
under a contract with NASA.
Based in part on data collected at Subaru Telescope and obtained from the
SMOKA, which is operated by the Astronomy Data Center, National Astronomical
Observatory of Japan.}

\author{Linhua Jiang\altaffilmark{1}, Kristian Finlator\altaffilmark{2,11},
Seth H. Cohen\altaffilmark{3}, Eiichi Egami\altaffilmark{4},
Rogier A. Windhorst\altaffilmark{3}, Xiaohui Fan\altaffilmark{4},
Romeel Dav\'{e}\altaffilmark{5}, Nobunari Kashikawa\altaffilmark{6}, 
Matthew Mechtley\altaffilmark{3}, Masami Ouchi\altaffilmark{7,8},
Kazuhiro Shimasaku\altaffilmark{9}, and Benjamin Cl{\'e}ment\altaffilmark{10}}

\altaffiltext{1}{Kavli Institute for Astronomy and Astrophysics, Peking 
   University, Beijing 100871, China; jiangKIAA@pku.edu.cn}
\altaffiltext{2}{Dark Cosmology Centre, Niels Bohr Institute, University of
   Copenhagen}
\altaffiltext{3}{School of Earth and Space Exploration, Arizona State
   University, Tempe, AZ 85287, USA}
\altaffiltext{4}{Steward Observatory, University of Arizona,
   933 North Cherry Avenue, Tucson, AZ 85721, USA}
\altaffiltext{5}{Physics Department, University of the Western Cape, 7535
   Bellville, Cape Town, South Africa}
\altaffiltext{6}{Optical and Infrared Astronomy Division, National
   Astronomical Observatory, Mitaka, Tokyo 181-8588, Japan}
\altaffiltext{7}{Institute for Cosmic Ray Research, The University of Tokyo,
   5-1-5 Kashiwanoha, Kashiwa, Chiba 277-8582, Japan}
\altaffiltext{8}{Kavli Institute for the Physics and Mathematics of the
   Universe, The University of Tokyo, 5-1-5 Kashiwanoha, Kashiwa, Chiba
   277-8583, Japan}
\altaffiltext{9}{Department of Astronomy, University of Tokyo, Hongo, Tokyo
   113-0033, Japan}
\altaffiltext{10}{Centre de Recherche Astrophysique de Lyon, Universite Lyon 
	1, 9 Avenue Charles Andre, F-69561 Saint Genis Laval Cedex, France}
\altaffiltext{11}{DARK fellow}

\begin{abstract}

We present a study of stellar populations in a sample of 
spectroscopically-confirmed Lyman-break galaxies (LBGs) and \lya\ emitters 
(LAEs) at $5.7<z<7$. These galaxies have deep optical and infrared images 
from Subaru, $HST$, and $Spitzer$/IRAC. We focus on a subset of 27 galaxies 
with IRAC detections, and characterize their stellar populations utilizing 
galaxy synthesis models based on the multi-band data and secure redshifts. By 
incorporating nebular emission estimated from the observed \lya\ flux, we are 
able to break the strong degeneracy of model spectra between young galaxies 
with prominent nebular emission and older galaxies with strong Balmer breaks.
The results show that our galaxies cover a wide range of ages from several to
a few hundred million years (Myr), and a wide range of stellar masses from 
$\sim$10$^8$ to $\sim$10$^{11}\,M_{\sun}$. These galaxies can be roughly 
divided into an `old' subsample and a `young' subsample. The `old' subsample 
consists of galaxies older than 100 Myr, with stellar masses higher than 
$10^9\,M_{\sun}$. The galaxies in the `young' subsample are younger than 
$\sim$30 Myr, with masses ranging between 
$\sim$10$^8$ and $\sim$ $3\times10^9\,M_{\sun}$. 
Both subsamples display a correlation between stellar mass and star-formation 
rate (SFR), but with very different normalizations. The average specific SFR
(sSFR, derived from a smoothly rising star-formation history) of the `old' 
subsample is 3--4 Gyr$^{-1}$, consistent with previous studies of `normal' 
star-forming galaxies at $z\ge6$. The average sSFR of the `young' subsample is 
an order of magnitude higher, likely due to starburst activity.
Our results also indicate little or no dust extinction in the majority
of the galaxies, as already suggested by their steep rest-frame UV slopes.
Finally, LAEs and LBGs with strong \lya\ emission are indistinguishable in 
terms of age, stellar mass, and SFR.

\end{abstract}


\section{INTRODUCTION}

Star-forming galaxies at $z\ge6$ are natural tools to study the early galaxy 
formation and explore the history of cosmic reionization. In recent years, with 
the advances of instrumentation on the {\it Hubble Space Telescope} ($HST$) 
and large ground-based telescopes, the number of galaxies found at $z\ge6$ has 
increased dramatically. The majority of these galaxies were photometrically 
selected Lyman-break galaxies (LBGs) using the dropout technique 
\citep[e.g.,][]{oes10,yan12,ell13,wil13,bou15,lap15}. 
Some of them, among the brightest in the optical and near-IR, have been 
spectroscopically confirmed with deep observations 
\citep[e.g.,][]{jia11,cur12,ono12,schenker12,fin13,pen14,oes15,wat15}.
A complementary way to find $z\ge6$ galaxies is the narrow-band (or \lya)
technique. This technique has a high success rate of spectroscopic 
confirmation. More than 200 \lya\ emitters (LAEs) at $z\ge6$ have been 
spectroscopically identified \citep[e.g.,][]{shi06,hu10,ouc10,kas11,hen12},
including several at $z\ge7$ \citep[e.g.,][]{iye06,rho12,shi12,kon14}.
Many properties of high-redshift LAEs can be used to probe cosmic 
reionization \citep[e.g.,][]{sil13,treu13,cai14,dij14,jen14,mom14}.

Meanwhile, the physical properties of $z\ge6$ galaxies are also being 
investigated. At $z\ge6$, the rest-frame UV/optical light moves to the IR 
range. Therefore, IR observations, including the $HST$ near-IR and 
{\it Spitzer Space Telescope} ($Spitzer$) mid-IR observations, are essential 
to measure the properties of these galaxies. While properties such as the 
rest-frame UV slope and galaxy morphology can be directly measured from 
optical and near-IR images \citep[e.g.,][]{mcl11,dun13,bou14,cur15,kaw15},
detailed physical properties such as age and stellar mass have to come from 
SED modeling of stellar populations based on the combination of optical, 
near-IR, and mid-IR data \citep{con13}. 
The optical data and $HST$ near-IR data measure the slope of the 
rest-frame UV spectrum, and constrain the properties of young stellar 
populations. $Spitzer$ IRAC provides mid-IR photometry. When combined with 
$HST$ near-IR data, it measures the amplitude of the Balmer break and 
constrains the properties of older stellar populations. Soon after the launch 
of $Spitzer$, it was found that IRAC is sensitive enough to directly detect
luminous $z\ge6$ LBGs \citep[e.g.,][]{ega05,eyl05,yan05}. In these early 
results, $z\simeq6$ LBGs showed strong IRAC detections, suggesting the 
existence of established massive stellar populations in their galaxies. 
Later studies then found that most galaxies were actually not detected in 
moderately deep IRAC images, meaning that they were considerably younger and 
less massive \citep[e.g.,][]{yan06,eyl07,pir07}. Extensive studies are now 
being carried out with various galaxy samples, and a diversity of physical 
properties are found \citep[e.g.,][]{gon10,sch10,mcl11,lab13,cur13,deb14}. 
These studies are mostly based on photometrically-selected samples, and their 
galaxies are not spectroscopically confirmed. 

This paper is the third in a series presenting the physical properties of a 
large sample of spectroscopically confirmed galaxies at $z\ge6$. In the first 
paper of the series \citep[hereafter Paper I]{jia13a}, we presented deep $HST$ 
and $Spitzer$ observations of 67 spectroscopically-confirmed galaxies at 
$z\ge6$. 
The sample is the largest collection of spectroscopically confirmed galaxies 
in this redshift range, including 51 LAEs and 16 LBGs. We measured basic 
properties of the rest-frame UV continuum and \lya\ emission in these galaxies.
In the second paper of the series \citep[hereafter Paper II]{jia13b}, we
carried out a structural and morphological study of these galaxies. In this 
third paper we will fit and model the SEDs of the galaxies, and derive 
physical parameters such as age, stellar mass, and dust extinction. This is 
particularly important for high-redshift LAEs. 
Some early observations \citep[e.g.,][]{mal02,pir07} found that $z\ge4.5$ LAEs 
consist of young stellar populations with low stellar masses (compared to 
LBGs). However, this could be due to selection effects \citep[e.g.,][]{day12}.
For LAEs at $z\ge5.7$, we currently have very little knowledge of their 
stellar populations. This is because almost all the known LAEs were discovered 
by ground-based telescopes, and do not have sufficiently
deep infrared observations. Our paper includes a large sample of LAEs. This 
allows us, for the first time, to systematically study stellar populations in 
$z>5.7$ LAEs.

As mentioned in Paper I, the spectroscopic redshifts of this sample have great 
advantages for measuring physical properties of high-redshift galaxies. Secure 
redshifts remove one critical free parameter in the SED modeling. A model 
spectrum of a bright $z\ge6$ galaxy is usually derived from 4--5 broad-band 
photometric points, e.g., 1 optical band, 2 $HST$ bands, 1--2 $Spitzer$ IRAC 
bands. Given the limited 
degrees of freedom, a spectroscopic redshift will significantly improve SED 
modeling, especially when gaseous or nebular emission is considered. 
It has been clear that strong nebular emission widely exists in high-redshift 
galaxies. For example, \citet{sta13} found that \ha\ contributes more than 
30\% of the IRAC 1 flux in $4<z<5$ galaxies. In the analysis of a large LBG 
sample at $3<z<6$, \citet{deb14} estimated that about 60--70\% of the LBGs 
show prominent nebular emission lines. At $z\ge6$, strong nebular lines such 
as [O\,{\sc iii}] $\lambda$5007, \hb, and \ha\ enter the IRAC 1 and 2 bands 
(3.6 and 4.5 $\mu$m), 
which significantly affects the measurements of stellar populations
\citep[e.g.,][]{rob10,sch10,finlator11,sta13,deb14}. Photometric redshifts 
with large uncertainties may place these nebular lines in the wrong filter
bands during SED fitting. With spectroscopic redshifts, the predicted 
observed wavelengths of nebular lines are securely known.

The structure of the paper is as follows. In Section 2 we briefly review our 
galaxy sample and the available data. In Section 3 we perform SED modeling of
the galaxies using evolutionary synthesis models. We then present the derived 
stellar populations of the galaxies in Section 4, and discuss the results
in Section 5. We summarize the paper in Section 6. Throughout the paper we 
adopt a $\Lambda$-dominated flat cosmology with $H_0=70$ 
km s$^{-1}$ Mpc$^{-1}$, $\Omega_{m}=0.3$, and $\Omega_{\Lambda}=0.7$.
All magnitudes are on the AB system \citep{oke83}.

\section{GALAXY SAMPLE AND DATA}

Our galaxy sample consists of 67 spectroscopically confirmed galaxies at 
$z\ge6$, including 62 galaxies in the Subaru Deep Field \citep[SDF;][]{kas04} 
and 5 galaxies in the Subaru XMM-Newton Deep Survey field 
\citep[SXDS;][]{fur08}. The SDF sample has 22 LAEs at $z\simeq5.7$,
25 LAEs at $z\simeq6.5$, one LAE at $z=6.96$, and 14 LBGs at $5.9\le z\le6.5$.
The SXDS sample contains 3 LAEs at $z\simeq6.5$ and 2 LBGs at $z\simeq6$.
All the LAEs at $z\simeq5.7$ and 6.5 have a relatively uniform magnitude limit 
of 26 mag in the narrow bands NB816 and NB921, and thus make a well-defined 
sample. The LBGs were selected with different criteria, and have 
rather inhomogeneous depth, so they are not a statistically complete sample.
The details of our galaxy sample can be found in Section 2 of Paper I. 
In the next two subsections we will briefly describe the data used for
this paper.

\subsection{Optical and Near-IR Imaging Data}

The optical imaging data for the two fields SDF and SXDS were obtained with
Subaru Suprime-Cam \citep{kas04,fur08}. They consists of images in a series of
broad and narrow bands. In Paper I, we produced a set of stacked images in six
broad bands ($BVRi'z'y$) and three narrow bands (NB816, NB921, and NB973), by
including all available data in the archive. Our stacked images have great
depth with excellent PSF FWHM of $0\farcs6-0\farcs7$. In particular, the
total integration time of the SDF $z'$ and $y$ band images is 29 hr and 24 hr,
corresponding to a depth of 27.1 mag and 26.2 mag, respectively ($5\sigma$
detection in a $2\arcsec$ diameter aperture). For the galaxies at
$z<6$, these two bands do not cover the \lya\ emission line, so they provide
two important photometric points for SED modeling.

We obtained near-IR imaging data for the SDF galaxies in three $HST$ GO
programs (11149: PI E. Egami; 12329 and 12616: PI L. Jiang).
The $HST$ observations were made with NICMOS and WFC3. The majority
of the galaxies were observed with WFC3 in the F125W (hereafter $J_{125}$) and
F160W (hereafter $H_{160}$) bands. The typical integration time was two $HST$
orbits (roughly 5400 sec) per band, leading to a depth of $\sim27.4$ mag
($5\sigma$ detection) in $J_{125}$ and a depth of $\sim27.1$ mag in $H_{160}$
\citep[see also][]{win11}.
The remaining several SDF galaxies were observed with NICMOS in
the F110W (hereafter $J_{110}$) and $H_{160}$ bands. The typical integration
time was also two $HST$ orbits, and the depth in the two bands are $\sim26.4$
mag and $\sim26.1$ mag, respectively. The five SXDS galaxies were covered by
the UKIDSS Ultra-Deep Survey (UDS). Their $HST$ WFC3 near-IR data were
obtained from the Cosmic Assembly Near-infrared Deep Extragalactic Legacy
Survey \citep[CANDELS;][]{gro11,koe11}. The exposure depth of the CANDELS UDS
data is 1900 sec in $J_{125}$ and 3300 sec in $H_{160}$, slightly shallower
than our WFC3 data for the SDF.
All the optical and near-IR photometry is listed in Table 1 of Paper I.

Our galaxies represent the most luminous galaxies at $z\ge6$, in terms of
\lya\ luminosity (for LAEs) or UV continuum luminosity (for LBGs). They cover
the brightest UV luminosity range of $M_{1500}<-19.5$ mag, so the majority of
them were detected ($>5\sigma$) in the near-IR images. In Paper I and Paper II
we have shown that these galaxies have steep UV continuum slopes $\beta$
with a median value of $\beta=-2.3$.
They have moderately strong rest-frame \lya\
equivalent width (EW) in the range $\sim$10 to $\sim$200 \AA. Their
star-formation rates (SFRs) are moderate from a few to a few tens solar masses
per year. These galaxies also exhibit a wide range of rest-frame UV continuum
morphology in the $HST$ images, from compact features to multiple component
systems. In this paper we will measure stellar populations in these galaxies.

\subsection{$Spitzer$ Mid-IR Imaging Data}

\begin{deluxetable*}{ccccccccc}
%\tabletypesize{\scriptsize}
\tablecaption{Mid-IR Photometry of the Galaxies in Our Sample}
%\tablewidth{0pt}
\tablehead{\colhead{No.} & \colhead{R.A. (J2000)} & \colhead{Decl. (J2000)}
	& \colhead{Redshift} & \colhead{IRAC 1} & \colhead{IRAC 2}}
\startdata
  2 & 13:23:54.601 & +27:24:12.72 & 5.654 &        $>$26.85 &        $\ldots$ \\
  3 & 13:24:16.468 & +27:19:07.65 & 5.665 &  25.09$\pm$0.21 &  25.39$\pm$0.33 \\
  4 & 13:24:32.885 & +27:30:08.82 & 5.671 &  24.78$\pm$0.22 &        $\ldots$ \\
  5 & 13:24:11.887 & +27:41:31.81 & 5.681 &        $>$26.64 &        $\ldots$ \\
 10 & 13:24:33.097 & +27:29:38.58 & 5.696 &        $>$26.35 &        $\ldots$ \\
 15 & 13:24:23.705 & +27:33:24.82 & 5.710 &  24.04$\pm$0.07 &  24.29$\pm$0.11 \\
 17 & 13:23:44.747 & +27:24:26.81 & 5.716 &        $>$26.56 &        $\ldots$ \\
 20 & 13:24:40.527 & +27:13:57.91 & 5.724 &  25.20$\pm$0.30 &        $\ldots$ \\
 21 & 13:24:30.633 & +27:29:34.61 & 5.738 &        $>$26.16 &        $\ldots$ \\
 22 & 13:24:41.264 & +27:26:49.09 & 5.743 &        $>$26.43 &        $\ldots$ \\
 23 & 13:24:18.450 & +27:16:32.56 & 5.922 &  25.11$\pm$0.16 &        $\ldots$ \\
 24 & 13:25:19.463 & +27:18:28.51 & 6.002 &  25.15$\pm$0.22 &        $\ldots$ \\
 25 & 13:24:26.559 & +27:15:59.72 & 6.032 &  24.30$\pm$0.07 &  24.47$\pm$0.13 \\
 27 & 13:24:10.766 & +27:19:03.95 & 6.040 &  25.70$\pm$0.30 &        $\ldots$ \\
 28 & 13:24:42.452 & +27:24:23.35 & 6.042 &  25.59$\pm$0.39 &        $\ldots$ \\
 29 & 13:24:05.895 & +27:18:37.72 & 6.049 &        $>$26.56 &        $\ldots$ \\
 30 & 13:24:00.301 & +27:32:37.95 & 6.062 &  25.50$\pm$0.35 &        $\ldots$ \\
 31 & 13:23:45.632 & +27:17:00.53 & 6.112 &  25.06$\pm$0.20 &        $\ldots$ \\
 33 & 13:24:20.628 & +27:16:40.47 & 6.269 &        $>$26.62 &        $\ldots$ \\
 34 & 13:23:45.757 & +27:32:51.30 & 6.315 &  23.82$\pm$0.09 &        $\ldots$ \\
 35 & 13:24:40.643 & +27:36:06.94 & 6.332 &  23.23$\pm$0.05 &  23.74$\pm$0.11 \\
 36 & 13:23:45.937 & +27:25:18.06 & 6.482 &  25.00$\pm$0.19 &  25.20$\pm$0.28 \\
 37 & 13:24:18.416 & +27:33:44.97 & 6.508 &        $>$26.61 &        $\ldots$ \\
 39 & 13:23:43.190 & +27:24:52.04 & 6.534 &        $>$26.43 &        $\ldots$ \\
 40 & 13:24:55.772 & +27:40:15.31 & 6.534 &        $>$26.59 &        $\ldots$ \\
 43 & 13:23:53.054 & +27:16:30.75 & 6.542 &  25.30$\pm$0.24 &        $\ldots$ \\
 44 & 13:24:15.678 & +27:30:57.79 & 6.543 &  23.77$\pm$0.05 &  24.02$\pm$0.09 \\
 45 & 13:24:40.239 & +27:25:53.11 & 6.544 &        $>$26.39 &        $\ldots$ \\
 46 & 13:23:52.680 & +27:16:21.76 & 6.545 &        $>$26.38 &        $\ldots$ \\
 47 & 13:24:10.817 & +27:19:28.08 & 6.547 &  23.70$\pm$0.05 &  23.74$\pm$0.09 \\
 48 & 13:23:48.922 & +27:15:30.33 & 6.548 &        $>$26.35 &        $\ldots$ \\
 49 & 13:24:17.909 & +27:17:45.94 & 6.548 &  25.33$\pm$0.24 &        $\ldots$ \\
 50 & 13:23:44.896 & +27:31:44.90 & 6.550 &  24.56$\pm$0.16 &        $\ldots$ \\
 52 & 13:24:35.005 & +27:39:57.43 & 6.554 &        $>$26.48 &        $\ldots$ \\
 54 & 13:24:08.313 & +27:15:43.49 & 6.556 &  25.14$\pm$0.25 &  25.21$\pm$0.36 \\
 58 & 13:24:43.427 & +27:26:32.62 & 6.583 &  25.44$\pm$0.26 &  24.72$\pm$0.23 \\
 61 & 13:25:22.291 & +27:35:19.95 & 6.599 &  24.08$\pm$0.07 &  24.50$\pm$0.18 \\
 62 & 13:23:59.766 & +27:24:55.75 & 6.964 &  24.84$\pm$0.16 &  25.17$\pm$0.21 \\
 63 & 02:18:00.899 & -05:11:37.69 & 6.023 &  24.49$\pm$0.12 &  24.97$\pm$0.25 \\
 64 & 02:17:35.337 & -05:10:32.50 & 6.116 &  24.90$\pm$0.11 &        $\ldots$ \\
 66 & 02:18:20.701 & -05:11:09.89 & 6.575 &  25.90$\pm$0.32 &        $\ldots$ \\
 67 & 02:17:57.585 & -05:08:44.72 & 6.595 &  23.72$\pm$0.04 &  24.25$\pm$0.08 \\
\enddata
\tablecomments{The sequence numbers of the galaxies in Column 1 correspond to
the numbers in Column 1 of Table 1 in Paper I.}
\end{deluxetable*}

Our $Spitzer$ IRAC imaging data for the SDF were obtained from two GO 
programs 40026 (PI: E. Egami) and 70094 (PI: L. Jiang). Program 40026 was 
carried out during the $Spitzer$ cryogenic phase, and the other one was 
carried out in the Warm Mission phase. The two programs imaged roughly 70\% of 
the SDF to a depth of 3--7 hours. %($\sim$25.5 AB mag). 
They covered all 62 SDF galaxies in our sample with IRAC channel 1 (3.6 
$\mu$m), and 51 (out of 62) galaxies with IRAC channel 2 (4.5 $\mu$m). 
The IRAC data were reduced with the $Spitzer$ 
Science Center (SSC) pipeline {\tt MOPEX}. The details of the $Spitzer$ 
observations and IRAC data reduction were described in Paper I. 
The final co-added images have a pixel size of $0\farcs6$, roughly a half of 
the IRAC native pixel scale. 

The IRAC images for the SXDS galaxies were obtained from the Spitzer Extended 
Deep Survey \citep[SEDS;][]{ashby13}. SEDS is a very deep imaging survey in 
the IRAC 1 and 2 bands over five well-studied fields, including the UDS.
With an integration time of 12 hr per pointing, SEDS reaches 26 mag in IRAC 1.
The SEDS co-added images also have a pixel scale of $0\farcs6$. The IRAC 
thumbnail images of all galaxies are shown in Figure 12 of Paper I.

The IRAC mid-IR photometry is complicated by source confusion. In deep IRAC 
images, faint galaxies are often blended with nearby neighbors, so reliable
photometry usually requires proper deblending and removal of neighbors. We 
performed source deblending using {\tt iGALFIT} \citep{ryan11}, an interactive 
tool to run {\tt GALFIT} \citep{pen02}. For each galaxy in our sample, the 
basic procedure is as follows. % We first cut a small thumbnail image centered
%on this galaxy from the co-added IRAC image. 
We first modeled its bright neighbors with
{\tt iGALFIT}. The model neighbors were convolved with a PSF image, and 
were subtracted from the original image. The PSF image was constructed from
a number of bright (but unsaturated) point sources. We then carried
out aperture photometry for this galaxy on the residual image. We used a 3 
pixel ($1\farcs8$) aperture radius, and computed its background in an annulus 
from 5 to 10 pixels. Finally we applied an aperture correction (roughly 0.4 
mag), which was measured from the PSF image. If a galaxy was isolated from
any bright neighboring objects, we performed aperture photometry directly, 
without removing neighbors.

The results of the above IRAC photometry are shown in Table 1, where we list 
the magnitudes (or $2\sigma$ upper limits) and errors for 42 (out of 67) 
galaxies in our sample. We have discarded the galaxies that have less than two
broad-band photometric points in the optical and near-IR (too few data points
for SED modeling). These discarded galaxies are all very faint in terms of 
their rest-frame UV continuum emission. They were barely (or not) detected in 
the $HST$ $J_{125}$ (or $J_{110}$) band, the deepest band that we have. This 
virtually puts a magnitude limit on our sample: galaxies fainter than 
$J_{125}\sim27.2$ mag were discarded. 
In Table 1, we have also discarded the galaxies that are heavily blended with 
(or completely covered by) much brighter neighbors in the IRAC images. In 
these cases, the IRAC photometry of the galaxies is not reliable. 
In Table 1, the sequence numbers of the galaxies in Column 1 correspond to
the numbers in Column 1 of Table 1 in Paper I. Column 2 lists the redshifts.
Columns 3 and 4 list the aperture photometry in IRAC 1 and 2 channels.
If a galaxy is not detected in IRAC 1, a $2\sigma$ upper limit is given.
We do not give upper limits for IRAC 2. Our IRAC 2 data are often 
significantly shallower than the IRAC 1 data, thus the inclusion of the IRAC 2
upper limits put no significant constraints on SED modeling.

\section{SED MODELING}

In this section we perform SED modeling using the {\tt GALEV} evolutionary
synthesis models \citep{kot09}. The {\tt GALEV} models are similar to other 
evolutionary synthesis models such as BC03 \citep{bc03} and STARBURST99 
\citep{lei99}. One distinct feature of {\tt GALEV} is that it provides an
option to include metallicity-dependent gaseous or nebular emission (both 
continuum and line emission). As we will see, nebular emission is critical
for the SED modeling of our galaxies.

For high-redshift galaxies, the quality of the SED fitting is usually 
dominated by data quality rather than the quality of synthesis models 
\citep[e.g.][]{pir12}, i.e., it is limited by the number of available 
photometric data points and photometric uncertainties. The galaxies in our 
sample, like $z\ge6$ galaxies in many other samples, have only 3--5 broad-band 
photometric points available, including 1--2 
ground-based optical points, 2 $HST$ near-IR points, and 1--2 IRAC mid-IR
points. In addition, these galaxies are faint, with relatively large 
photometric uncertainties particularly in the IRAC bands. Among 42 galaxies 
shown in Table 1, 27 galaxies were detected ($>3\sigma$) in our IRAC 1 
images, and 13 galaxies were also detected in the IRAC 2 images.
In order to account for the associated systematic uncertainties, we model our 
measurements under a broad range of assumptions, with the view that the true 
physical parameters of our galaxies lies somewhere within the range of 
possibilities that we consider.

\subsection{Stellar Populations with and without Nebular Emission}

\begin{deluxetable*}{llllllllllll}
%\tabletypesize{\scriptsize}
\tablecaption{Comparison between the SED-fitting Results from the NoEM Model 
and the GALEV-EM Model}
\tablewidth{0pt}
\tablehead{\colhead{No.} & \multicolumn{2}{c}{$M_{\ast}$ ($10^8 M_{\sun}$)} &
	\colhead{} & \multicolumn{2}{c}{Age (Myr)} & 
	\colhead{} & \multicolumn{2}{c}{$E(B-V)$} &
	\colhead{} & \multicolumn{2}{c}{$\chi_{r}^2$} \\
	\cline{2-3} \cline{5-6} \cline{8-9} \cline{11-12} \\
	\colhead{} & \colhead{NoEM} & \colhead{EM} &
	\colhead{} & \colhead{NoEM} & \colhead{EM} &
	\colhead{} & \colhead{NoEM} & \colhead{EM} &
	\colhead{} & \colhead{NoEM} & \colhead{EM} }
\startdata
 3 & $  62.4_{-  24.8}^{+  32.0}$ & $  18.1_{-   0.8}^{+   0.9}$ &  & $ 200_{- 104}^{+ 132}$ & $  40_{-  36}^{+ 152}$ &  & $ 0.04_{-0.04}^{+0.04}$ & $ 0.02_{-0.02}^{+0.06}$ &  &   1.4 &  1.5 \\
 4 & $ 114.0_{-  10.0}^{+  11.0}$ & $   3.5_{-   0.2}^{+   0.2}$ &  & $ 928_{- 316}^{+  68}$ & $   4_{-   0}^{+   8}$ &  & $ 0.00_{-0.00}^{+0.04}$ & $ 0.00_{-0.00}^{+0.04}$ &  &  23.6 & 18.8 \\
15 & $ 368.8_{-  16.6}^{+  35.6}$ & $  15.7_{-   0.7}^{+   0.7}$ &  & $ 976_{-  84}^{+  20}$ & $   4_{-   0}^{+   8}$ &  & $ 0.04_{-0.04}^{+0.04}$ & $ 0.08_{-0.04}^{+0.04}$ &  &   4.3 &  2.7 \\
20 & $ 166.0_{-  45.8}^{+  63.2}$ & $  44.1_{-  25.7}^{+  77.4}$ &  & $ 656_{- 404}^{+ 340}$ & $  88_{-  84}^{+ 420}$ &  & $ 0.14_{-0.06}^{+0.06}$ & $ 0.16_{-0.06}^{+0.06}$ &  &   0.2 &  0.2 \\
23 & $ 127.5_{-   5.7}^{+  12.3}$ & $   7.3_{-   0.3}^{+   0.3}$ &  & $ 988_{- 276}^{+   8}$ & $   8_{-   4}^{+  76}$ &  & $ 0.00_{-0.00}^{+0.04}$ & $ 0.04_{-0.04}^{+0.04}$ &  &   1.8 &  1.2 \\
24 & $  86.1_{-  11.1}^{+  12.8}$ & $   3.8_{-   0.2}^{+   0.2}$ &  & $ 632_{- 264}^{+ 364}$ & $   4_{-   0}^{+  16}$ &  & $ 0.00_{-0.00}^{+0.04}$ & $ 0.00_{-0.00}^{+0.06}$ &  &   2.4 &  0.8 \\
25 & $ 356.2_{-  16.0}^{+  34.4}$ & $  14.4_{-   0.6}^{+   0.7}$ &  & $ 992_{-  52}^{+   4}$ & $   4_{-   0}^{+   8}$ &  & $ 0.08_{-0.04}^{+0.04}$ & $ 0.12_{-0.04}^{+0.04}$ &  &  17.1 & 13.3 \\
27 & $  93.1_{-  31.6}^{+  35.4}$ & $   3.2_{-   0.3}^{+   0.3}$ &  & $ 988_{- 468}^{+   8}$ & $   4_{-   0}^{+  40}$ &  & $ 0.06_{-0.06}^{+0.06}$ & $ 0.08_{-0.08}^{+0.08}$ &  &   3.4 &  2.4 \\
28 & $  56.8_{-   9.6}^{+  41.9}$ & $   2.8_{-   0.4}^{+   0.3}$ &  & $ 536_{- 364}^{+ 460}$ & $   4_{-   0}^{+ 616}$ &  & $ 0.00_{-0.00}^{+0.08}$ & $ 0.00_{-0.00}^{+0.08}$ &  &   0.4 &  $<$0.1 \\
30 & $  86.9_{-  21.0}^{+  71.2}$ & $   3.0_{-   1.8}^{+   1.0}$ &  & $ 992_{- 448}^{+   4}$ & $   4_{-   0}^{+  24}$ &  & $ 0.02_{-0.02}^{+0.10}$ & $ 0.04_{-0.04}^{+0.08}$ &  &   3.7 &  2.7 \\
31 & $ 275.6_{-  93.5}^{+ 141.5}$ & $   8.3_{-   5.0}^{+   3.7}$ &  & $ 984_{- 384}^{+  12}$ & $   4_{-   0}^{+  16}$ &  & $ 0.16_{-0.08}^{+0.08}$ & $ 0.16_{-0.08}^{+0.08}$ &  &   1.8 &  1.0 \\
34 & $ 852.2_{-  38.4}^{+  82.2}$ & $  41.0_{-   1.8}^{+   1.9}$ &  & $ 992_{- 136}^{+   4}$ & $   8_{-   4}^{+  12}$ &  & $ 0.14_{-0.04}^{+0.04}$ & $ 0.16_{-0.04}^{+0.06}$ &  &   3.2 &  1.4 \\
35 & $3314.4_{- 291.6}^{+ 319.8}$ & $ 102.5_{-   4.6}^{+   4.8}$ &  & $ 976_{- 112}^{+  20}$ & $   4_{-   0}^{+   8}$ &  & $ 0.36_{-0.04}^{+0.04}$ & $ 0.36_{-0.04}^{+0.04}$ &  &  68.3 & 25.0 \\
36 & $ 157.7_{-  13.9}^{+  15.2}$ & $ 124.9_{-  11.0}^{+  12.1}$ &  & $ 992_{- 200}^{+   4}$ & $ 984_{- 392}^{+  12}$ &  & $ 0.02_{-0.02}^{+0.06}$ & $ 0.00_{-0.00}^{+0.06}$ &  &   3.5 &  2.9 \\
43 & $ 129.4_{-  26.6}^{+  41.2}$ & $   4.5_{-   0.4}^{+   0.4}$ &  & $ 992_{- 236}^{+   4}$ & $   4_{-   0}^{+  12}$ &  & $ 0.06_{-0.06}^{+0.08}$ & $ 0.08_{-0.08}^{+0.06}$ &  &   6.9 &  3.8 \\
44 & $ 983.9_{-  44.3}^{+  94.9}$ & $  68.0_{-   3.1}^{+   3.2}$ &  & $ 992_{-  96}^{+   4}$ & $  12_{-   8}^{+  12}$ &  & $ 0.16_{-0.04}^{+0.04}$ & $ 0.20_{-0.04}^{+0.04}$ &  &  10.0 &  2.3 \\
47 & $ 758.3_{- 127.6}^{+ 153.4}$ & $ 388.1_{-  79.8}^{+ 123.5}$ &  & $ 440_{-  80}^{+  60}$ & $ 312_{-  84}^{+  96}$ &  & $ 0.18_{-0.04}^{+0.04}$ & $ 0.14_{-0.04}^{+0.04}$ &  &   7.1 &  6.6 \\
49 & $ 207.8_{-  42.7}^{+ 121.6}$ & $   6.1_{-   3.7}^{+   0.6}$ &  & $ 996_{- 436}^{+   0}$ & $   4_{-   0}^{+ 992}$ &  & $ 0.12_{-0.08}^{+0.08}$ & $ 0.12_{-0.08}^{+0.08}$ &  &   0.4 &  0.1 \\
50 & $1223.8_{- 251.7}^{+ 389.5}$ & $  30.6_{-   6.3}^{+   7.9}$ &  & $ 996_{- 324}^{+   0}$ & $   4_{-   0}^{+  12}$ &  & $ 0.36_{-0.06}^{+0.08}$ & $ 0.34_{-0.06}^{+0.08}$ &  &   1.5 &  0.5 \\
54 & $ 144.7_{-  29.8}^{+  84.7}$ & $ 112.4_{-  14.5}^{+  42.8}$ &  & $ 992_{- 268}^{+   4}$ & $ 972_{- 412}^{+  24}$ &  & $ 0.02_{-0.02}^{+0.08}$ & $ 0.00_{-0.00}^{+0.06}$ &  &   1.9 &  1.6 \\
58 & $ 183.4_{-  56.5}^{+  81.7}$ & $ 178.8_{-  43.2}^{+  56.9}$ &  & $ 480_{- 216}^{+ 332}$ & $ 884_{- 380}^{+ 112}$ &  & $ 0.12_{-0.06}^{+0.06}$ & $ 0.06_{-0.06}^{+0.06}$ &  &   1.9 &  3.1 \\
61 & $ 875.4_{-  77.0}^{+ 129.7}$ & $  29.2_{-   1.3}^{+   1.4}$ &  & $ 984_{- 116}^{+  12}$ & $   4_{-   0}^{+   8}$ &  & $ 0.20_{-0.04}^{+0.04}$ & $ 0.22_{-0.04}^{+0.04}$ &  &  16.4 &  9.3 \\
62 & $ 194.5_{-  47.0}^{+  61.9}$ & $ 163.2_{-  45.0}^{+  62.1}$ &  & $ 756_{- 248}^{+ 240}$ & $ 656_{- 256}^{+ 304}$ &  & $ 0.04_{-0.04}^{+0.06}$ & $ 0.04_{-0.04}^{+0.06}$ &  &   1.6 &  1.5 \\
63 & $ 201.7_{-   9.1}^{+  19.5}$ & $   7.0_{-   0.3}^{+   0.3}$ &  & $ 988_{- 148}^{+   8}$ & $   4_{-   0}^{+   8}$ &  & $ 0.00_{-0.00}^{+0.04}$ & $ 0.02_{-0.02}^{+0.04}$ &  &   6.3 &  2.9 \\
64 & $ 167.1_{-  14.7}^{+  16.1}$ & $   5.0_{-   0.2}^{+   0.2}$ &  & $ 992_{- 200}^{+   4}$ & $   4_{-   0}^{+   8}$ &  & $ 0.00_{-0.00}^{+0.06}$ & $ 0.00_{-0.00}^{+0.06}$ &  &   6.5 &  3.8 \\
66 & $ 113.5_{-  53.9}^{+ 102.8}$ & $  23.0_{-  15.0}^{+  56.8}$ &  & $ 920_{- 636}^{+  76}$ & $  96_{-  92}^{+ 900}$ &  & $ 0.10_{-0.10}^{+0.10}$ & $ 0.12_{-0.10}^{+0.10}$ &  &  $<$0.1 &  $<$0.1 \\
67 & $ 774.8_{-  68.2}^{+  74.8}$ & $  27.1_{-   1.2}^{+   1.3}$ &  & $ 976_{- 108}^{+  20}$ & $   4_{-   0}^{+   8}$ &  & $ 0.10_{-0.04}^{+0.04}$ & $ 0.12_{-0.04}^{+0.04}$ &  &  27.2 &  3.3 \\
\enddata
\tablecomments{This table compares the SED-fitting results from the NoEM Model
and from the GALEV-EM Model with rSFH. The minimum and maximum values of age 
are 4 and 1000 Myr. When an age is close to the two limits, its errors are not 
reliable because of the lack of dynamic range. In addition, the fitting
results with very large or small $\chi_{r}^2$ ($\chi_{r}^2\gg1$ or
$\chi_{r}^2\ll1$) are not reliable.}
\end{deluxetable*}

Given the limited number of available photometric data points, we use as few 
free parameters as possible in our SED fitting. Compared to photometric 
samples, the advantage of our sample are the spectroscopic redshifts that 
remove one critical free parameter. The other major parameters for SED fitting 
are metallicity, dust extinction (or reddening), stellar mass, age, and star 
formation history (SFH). We adopt a Salpeter initial mass function (IMF) with 
a mass range of 0.1--100 $M_{\sun}$. 
For metallicity, the {\tt GALEV} models provide two 
options: chemically consistent treatment or fixed metallicity values. We 
choose to use the fixed metallicity values as other synthesis models do. 
There is a strong age-metallicity degeneracy at young ages (many of our 
galaxies are young), and our data are not sufficient enough to break this 
degeneracy. So we fix metallicity to be 
0.2 $Z_{\sun}$, which was suggested by the simulations in 
\citet[e.g.][]{finlator11}.

We use two representative SFHs, an exponentially declining SFH and a smoothly 
rising SFH. For the exponentially declining SFH $\sim {\rm exp}(-t/\tau)$, we 
fix the decline factor $\tau$ to be 200 million years (Myr) (to reduce the 
number of free parameters). The SFR of this SFH declines slowly at young ages 
(younger than $\sim200$ Myr) and is not sensitive to $\tau$ for $\tau>200$ 
Myr. At high redshift, a more realistic SFH is probably a smoothly rising SFH 
\citep[e.g.][]{finlator07,finlator11,papovich11}.
{\tt GALEV} has not included any rising SFHs yet, so we incorporate the 
smoothly rising SFH (SFR as a function of age) of \citet{finlator11} into the 
{\tt GALEV} models. Hereafter we denote the above exponentially declining SFH 
and smoothly rising SFH as `dSFH' and `rSFH', respectively. Our purpose of 
using two SFH models is to explore the possible ranges of physical parameters, 
rather than distinguishing one model from the other. We do not use more 
complex SFHs that usually introduce new free parameters. We do not consider 
the simple stellar population (SSP) or an instantaneous burst model. The SSP 
model is physically difficult to understand our $z\ge6$ galaxies. These 
galaxies have strong \lya\ emission, and presumably have strong nebular 
emission (we will discuss this later). But the nebular emission produced by 
the SSP drops rapidly in the first few Myr, as the instantaneous burst goes 
off.

From SED modeling, we mainly constrain three physical quantities, including 
dust reddening $E(B-V)$, stellar mass $M_{\ast}$, and age. We use the 
reddening law of \citet{cal00}, and allow $E(B-V)$ to range between 0.00 and 
0.50 in steps of 0.02. The age provided by the {\tt GALEV} models starts from 
4 Myr in steps of 4 Myr. We allow age to vary between 4 Myr and 1000 Myr, 
which approaches the age of the universe at $z\sim6$.
For $z\ge6$ galaxies with the limited number of photometric points, age is 
usually poorly constrained due to various degeneracies among parameters. 
On the contrary, mass is the amplitude of a model spectrum, and is thus 
thought to be more easily constrained \citep[e.g.][]{pap01,sha05,con13,mob15}.
In reality, the mass estimate depends on the mass-to-light (M/L) ratio, which 
in turn depends strongly on the age of stellar populations. The measurement of 
the M/L ratios in most normal-SED galaxies can be accurate to a level of 
$\sim0.3$ dex, if the uncertainty is dominated by systematics \citep{con13}. 
However, the M/L ratios can be very uncertain for certain types of galaxies, 
or galaxies in certain age ranges (see \citet{con13} for a review).

The {\tt GALEV} models provide options to include (or not) nebular line and 
continuum emission. We use both options for each individual model, and 
denote models with and without nebular/gaseous emission as `EM' and `NoEM' 
models, respectively. In the rest of the paper, we mostly discuss the 27 
galaxies that were detected in the IRAC 1 band (galaxy parameters cannot be 
properly constrained for our galaxies without IRAC 1 detections). 
We fit the models to the SEDs of the galaxies and derive the above parameters
by the minimum $\chi^2$ method. These results are shown in Table 2.
The 1$\sigma$ uncertainties quoted in the table are estimated in a standard
way, i.e., allowing all other parameters to vary 
until $\Delta \chi^2=1$. Figure 1 
compares the distributions of stellar mass, age, and $E(B-V)$ in the 27 
galaxies for different models and SFHs. The two columns of panels from left 
to right in Figure 1 correspond to the dSFH and the rSFH models, respectively. 
The grey filled histograms represent the EM models, and the black unfilled 
histograms represent the NoEM models.

\begin{figure}
\epsscale{1.1}
\plotone{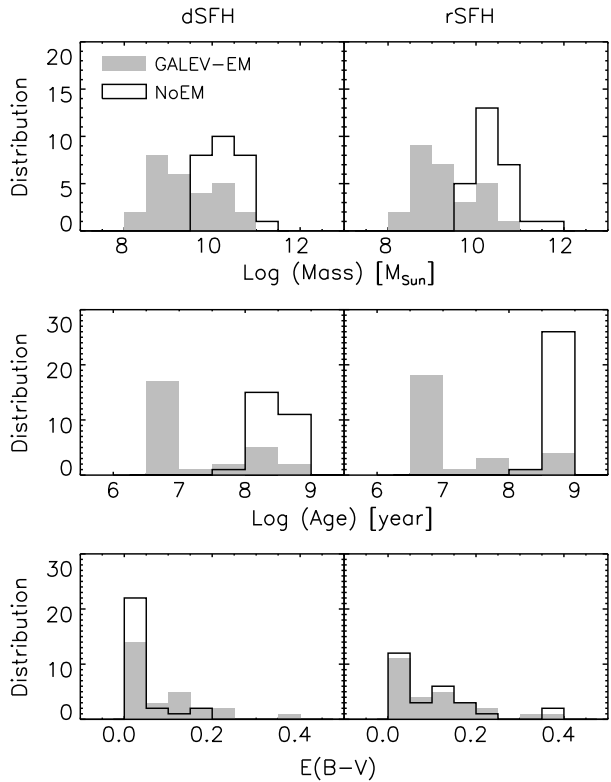}
\caption{Distributions of the stellar masses, ages, and $E(B-V)$ estimated
for the 27 galaxies with the IRAC 1 detections. The two columns of the panels
from the left to the right correspond to the declining SFH and the rising SFH,
respectively. The grey filled histograms represent the EM models, and the
black unfilled histograms represent the NoEM models. The NoEM and EM
models produce different stellar populations (old and massive versus young
and less massive) for the same galaxies. This is due to the strong degeneracy
between young galaxies with prominent nebular emission and old galaxies with
strong Balmer breaks.}
\end{figure}

Table 2 and Figure 1 show that the EM and NoEM models produce very
different stellar populations for the same galaxies. Stellar populations of
many galaxies from the EM models are very young, with ages of several Myr.
They also have relatively low stellar masses. On the contrary, stellar 
populations from the NoEM models are mostly older than a few hundred Myr.
They are usually very massive, with masses close to or higher than 
$10^{10}\,M_{\sun}$. Based on the minimum $\chi^2$ values, the quality of
the EM models is comparable to (in some cases marginally better than) that of
the NoEM models. The results are expected \citep[e.g.][]{sch09}, due to the 
the degeneracy between young galaxies with prominent nebular emission and old
galaxies with strong Balmer breaks.
This degeneracy is particularly strong for the galaxies in our sample for the 
following reasons. First, our galaxies are at $5.7<z<6.6$, so both IRAC 
1 and 2 bands cover some of the strongest emission lines such as \oiii, \hb, 
and \ha, etc. The second reason is that the IRAC 1 and 2 data have the largest 
photometric uncertainties compared to other bands; many galaxies were even not 
detected in the IRAC 2 band. Finally, for this redshift range, the 
wavelength range that IRAC 1 and 2 cover mimics the Balmer continuum/break
that the synthesis models rely on to constrain stellar populations.
The combination of these reasons make the EM and NoEM models
indistinguishable for our galaxies. 

\begin{figure}
\epsscale{1.1}
\plotone{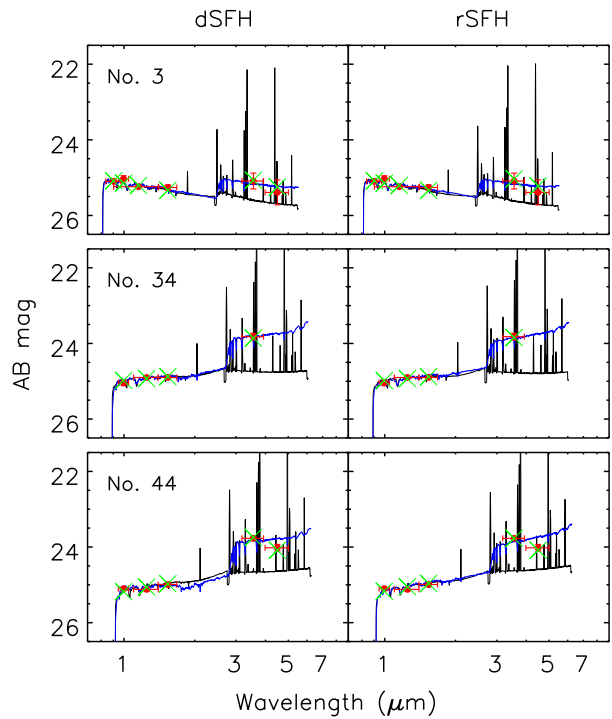}
\caption{Three examples of SED modeling illustrating the strong degeneracy
between young galaxies with prominent nebular emission and older galaxies
with strong Balmer breaks. The two columns of the panels correspond to
the declining SFH and the rising SFH. The red points with error
bars are the observed photometric data points. The horizontal errors indicate
the wavelength ranges of the filters. The black and blue profiles represent
the EM and NoEM models, respectively. The large green crosses represent
the photometric points predicted by the EM models.
The existence of emission lines in the
EM models exactly mimics the the Balmer continua/breaks in the NoEM models,
which significantly reduces the required ages and stellar masses
of stellar populations in the EM models.}
\end{figure}

Figure 2 illustrates the degeneracy mentioned above by the SED modeling of
three galaxies, one LAE at $z=5.664$ (No. 3), one LBG at $z=6.315$ (No. 34),
and one LAE at $z=6.543$ (No. 44). The two columns of the panels from left
to right correspond to the dSFH and the rSFH models, respectively.
The red points with error bars are the observed photometric data points. The
black and blue SED profiles represent the EM and NoEM models. Note that the
{\tt GALEV} models do not include the \lya\ emission line. IGM absorption
has been applied to the model spectra. We calculate the IGM absorption using
the method of \citet{fan01} and \citet{jia08}. Note that the \lya\ emission
or the IGM absorption does not affect our SED fitting, because we did not use 
the bands that cover \lya. The figure shows that the
existence of emission lines in the EM models mimics the Balmer continua/breaks
in the NoEM models, which significantly reduces the age and stellar mass
of stellar populations needed in the EM models.

\subsection{Nebular Emission Estimated from the \lya\ Line Emission}

\begin{deluxetable*}{llllllcccccccc}
%\tabletypesize{\scriptsize}
\tablecaption{SED-fitting Results with the \lya-EM models}
%\tablewidth{0pt}
\tablehead{\colhead{No.} & \multicolumn{2}{c}{$M_{\ast}$ ($10^8 M_{\sun}$)} &
	\colhead{} & \multicolumn{2}{c}{Age (Myr)} & 
	\colhead{} & \multicolumn{2}{c}{$E(B-V)$} &
	\colhead{} & \multicolumn{2}{c}{$\chi_{r}^2$} \\
	\cline{2-3} \cline{5-6} \cline{8-9} \cline{11-12} \\
\colhead{} & \colhead{dSFH} & \colhead{rSFH} &
\colhead{} & \colhead{dSFH} & \colhead{rSFH} &
\colhead{} & \colhead{dSFH} & \colhead{rSFH} &
\colhead{} & \colhead{dSFH} & \colhead{rSFH} }
\startdata
 3 & $   7.2_{-   0.3}^{+   0.3}$ & $   7.2_{-   0.3}^{+   0.3}$ &  & $   4_{-   0}^{+  12}$ & $   4_{-   0}^{+  20}$ &  & $ 0.04_{-0.04}^{+0.06}$ & $ 0.04_{-0.04}^{+0.06}$ &  &   1.5 &  1.4 \\
 4 & $   3.4_{-   0.1}^{+   0.2}$ & $   3.4_{-   0.1}^{+   0.2}$ &  & $   4_{-   0}^{+   8}$ & $   4_{-   0}^{+   8}$ &  & $ 0.00_{-0.00}^{+0.04}$ & $ 0.00_{-0.00}^{+0.04}$ &  &  18.1 & 18.1 \\
15 & $ 127.3_{-  11.2}^{+  12.3}$ & $  21.1_{-   0.9}^{+   1.0}$ &  & $ 168_{-  24}^{+  32}$ & $   8_{-   4}^{+  12}$ &  & $ 0.00_{-0.00}^{+0.04}$ & $ 0.08_{-0.04}^{+0.04}$ &  &   2.8 &  2.9 \\
20 & $  29.5_{-   9.1}^{+  13.1}$ & $  32.8_{-  10.1}^{+  21.6}$ &  & $ 148_{- 128}^{+ 164}$ & $ 492_{- 468}^{+ 504}$ &  & $ 0.02_{-0.02}^{+0.06}$ & $ 0.02_{-0.02}^{+0.08}$ &  &   1.5 &  1.6 \\
23 & $  46.8_{-   6.0}^{+   9.5}$ & $  57.9_{-   9.8}^{+  11.7}$ &  & $ 124_{-  48}^{+  52}$ & $ 444_{- 208}^{+ 336}$ &  & $ 0.00_{-0.00}^{+0.04}$ & $ 0.00_{-0.00}^{+0.04}$ &  &   1.7 &  1.5 \\
24 & $   3.8_{-   0.2}^{+   0.2}$ & $   3.8_{-   0.2}^{+   0.2}$ &  & $   4_{-   0}^{+  20}$ & $   4_{-   0}^{+  28}$ &  & $ 0.00_{-0.00}^{+0.06}$ & $ 0.00_{-0.00}^{+0.06}$ &  &   0.8 &  0.8 \\
25 & $  20.9_{-   0.9}^{+   1.0}$ & $  66.5_{-   5.8}^{+   6.4}$ &  & $  12_{-   8}^{+  20}$ & $ 956_{- 184}^{+  40}$ &  & $ 0.10_{-0.04}^{+0.04}$ & $ 0.02_{-0.02}^{+0.04}$ &  &  13.2 & 12.8 \\
27 & $  12.6_{-   1.6}^{+   6.5}$ & $  17.4_{-   4.2}^{+  14.2}$ &  & $  88_{-  84}^{+ 160}$ & $ 324_{- 320}^{+ 672}$ &  & $ 0.00_{-0.00}^{+0.08}$ & $ 0.00_{-0.00}^{+0.06}$ &  &   2.2 &  2.2 \\
28 & $   2.7_{-   0.2}^{+   0.3}$ & $   2.7_{-   0.2}^{+   0.3}$ &  & $   4_{-   0}^{+  20}$ & $   4_{-   0}^{+  28}$ &  & $ 0.00_{-0.00}^{+0.04}$ & $ 0.00_{-0.00}^{+0.04}$ &  &   1.9 &  1.9 \\
30 & $  43.4_{-   8.9}^{+  28.6}$ & $  26.8_{-   4.5}^{+  15.7}$ &  & $ 192_{- 136}^{+ 184}$ & $ 912_{- 728}^{+  84}$ &  & $ 0.00_{-0.00}^{+0.08}$ & $ 0.00_{-0.00}^{+0.08}$ &  &   3.2 &  3.1 \\
31 & $   2.4_{-   1.7}^{+   0.6}$ & $   2.4_{-   2.2}^{+   0.6}$ &  & $   4_{-   0}^{+ 168}$ & $   4_{-   0}^{+ 668}$ &  & $ 0.04_{-0.04}^{+0.08}$ & $ 0.04_{-0.04}^{+0.08}$ &  &   0.2 &  0.2 \\
34 & $ 640.7_{- 107.8}^{+ 129.6}$ & $ 250.0_{-  11.2}^{+  24.1}$ &  & $ 356_{-  44}^{+  44}$ & $ 992_{- 204}^{+   4}$ &  & $ 0.06_{-0.04}^{+0.04}$ & $ 0.12_{-0.04}^{+0.04}$ &  &   1.7 &  2.1 \\
35 & $  54.2_{-   2.4}^{+   2.5}$ & $  81.1_{-   3.7}^{+   3.8}$ &  & $ 592_{-   8}^{+  68}$ & $   4_{-   0}^{+   8}$ &  & $ 0.00_{-0.00}^{+0.04}$ & $ 0.34_{-0.06}^{+0.04}$ &  &   1.1 &  2.8 \\
36 & $ 109.8_{-  18.5}^{+  28.4}$ & $  46.4_{-   4.1}^{+   4.5}$ &  & $ 240_{-  60}^{+  72}$ & $ 988_{- 236}^{+   8}$ &  & $ 0.00_{-0.00}^{+0.06}$ & $ 0.00_{-0.00}^{+0.06}$ &  &   3.3 &  3.1 \\
43 & $  76.3_{-  15.7}^{+  34.0}$ & $  30.8_{-   7.4}^{+   9.8}$ &  & $ 284_{- 120}^{+ 128}$ & $ 972_{- 528}^{+  24}$ &  & $ 0.00_{-0.00}^{+0.06}$ & $ 0.02_{-0.02}^{+0.06}$ &  &   4.4 &  4.4 \\
44 & $ 331.1_{-  29.1}^{+  31.9}$ & $ 172.9_{-  15.2}^{+  16.7}$ &  & $ 260_{-  32}^{+  32}$ & $ 872_{- 160}^{+ 124}$ &  & $ 0.06_{-0.04}^{+0.04}$ & $ 0.08_{-0.04}^{+0.04}$ &  &   1.1 &  1.0 \\
47 & $ 368.7_{-  16.6}^{+  17.4}$ & $ 391.1_{- 107.8}^{+ 101.3}$ &  & $  76_{-  20}^{+  20}$ & $ 184_{-  40}^{+  68}$ &  & $ 0.18_{-0.04}^{+0.04}$ & $ 0.18_{-0.04}^{+0.04}$ &  &   5.8 &  5.9 \\
49 & $  59.3_{-  21.9}^{+  30.5}$ & $  35.6_{-  16.9}^{+  35.4}$ &  & $ 196_{- 148}^{+ 144}$ & $ 212_{- 208}^{+ 748}$ &  & $ 0.04_{-0.04}^{+0.08}$ & $ 0.08_{-0.08}^{+0.08}$ &  &   0.1 &  0.1 \\
50 & $ 173.5_{-  97.8}^{+  77.3}$ & $   9.0_{-   1.9}^{+   0.9}$ &  & $ 528_{-  72}^{+  40}$ & $   4_{-   0}^{+ 548}$ &  & $ 0.06_{-0.06}^{+0.06}$ & $ 0.22_{-0.08}^{+0.06}$ &  &   $<$0.1 &  $<$0.1 \\
54 & $  36.5_{-   6.2}^{+  13.9}$ & $  51.5_{-  10.6}^{+  23.0}$ &  & $ 112_{-  84}^{+ 100}$ & $ 460_{- 392}^{+ 524}$ &  & $ 0.00_{-0.00}^{+0.06}$ & $ 0.00_{-0.00}^{+0.06}$ &  &   1.9 &  1.8 \\
58 & $ 138.7_{-  33.5}^{+  44.1}$ & $  61.6_{-  14.9}^{+  19.6}$ &  & $ 268_{-  76}^{+  80}$ & $ 960_{- 384}^{+  36}$ &  & $ 0.02_{-0.02}^{+0.06}$ & $ 0.04_{-0.04}^{+0.06}$ &  &   3.7 &  3.6 \\
61 & $ 357.6_{-  31.5}^{+  53.0}$ & $  66.3_{-   3.0}^{+   3.1}$ &  & $ 516_{-  24}^{+  24}$ & $  32_{-  28}^{+  80}$ &  & $ 0.00_{-0.00}^{+0.04}$ & $ 0.20_{-0.04}^{+0.04}$ &  &   8.2 &  9.5 \\
62 & $ 128.0_{-  21.5}^{+  33.1}$ & $ 115.0_{-  31.7}^{+  51.2}$ &  & $ 228_{-  64}^{+  56}$ & $ 472_{- 216}^{+ 316}$ &  & $ 0.00_{-0.00}^{+0.06}$ & $ 0.04_{-0.04}^{+0.06}$ &  &   1.2 &  1.2 \\
63 & $   9.8_{-   0.4}^{+   0.5}$ & $   9.4_{-   0.4}^{+   0.5}$ &  & $   8_{-   4}^{+  20}$ & $   8_{-   4}^{+  36}$ &  & $ 0.02_{-0.02}^{+0.04}$ & $ 0.02_{-0.02}^{+0.04}$ &  &   3.5 &  3.5 \\
64 & $  35.2_{-   3.1}^{+   9.1}$ & $  41.3_{-  11.4}^{+  30.5}$ &  & $  72_{-  44}^{+  68}$ & $ 200_{- 136}^{+ 260}$ &  & $ 0.00_{-0.00}^{+0.06}$ & $ 0.00_{-0.00}^{+0.06}$ &  &   5.5 &  5.4 \\
66 & $   1.4_{-   0.8}^{+   0.7}$ & $   1.4_{-   1.2}^{+   0.7}$ &  & $   4_{-   0}^{+ 244}$ & $   4_{-   0}^{+ 992}$ &  & $ 0.00_{-0.00}^{+0.08}$ & $ 0.00_{-0.00}^{+0.08}$ &  &   0.3 &  0.3 \\
67 & $  32.0_{-   1.4}^{+   1.5}$ & $  32.0_{-   1.4}^{+   1.5}$ &  & $   4_{-   0}^{+   8}$ & $   4_{-   0}^{+   8}$ &  & $ 0.14_{-0.04}^{+0.04}$ & $ 0.14_{-0.04}^{+0.04}$ &  &   3.2 &  3.2 \\
\enddata
\tablecomments{This table shows the SED-fitting results using the \lya-EM
models, i.e., models with nebular emission estimated from the observed \lya\
line flux (Section 3.2). The minimum and maximum values of age are 4 and 1000 
Myr. When an age is close to the two limits, its 
errors are not reliable because of the lack of dynamic range.}
\end{deluxetable*}

With ongoing star formation in the models with rising and declining SFHs,
nebular emission is naturally expected in our galaxies. In particular, we have
seen strong \lya\ emission lines in these galaxies. However, the actual
strength of nebular emission, including continuum and line emission, is
unknown. In the {\tt GALEV} models that we use, the nebular
emission is metallicity dependent, but its relative strength to the continuum
is fixed in individual model galaxies. In other words, at any given
age and metallicity for the same SFH, all galaxies have the same strength of
nebular emission. This is a model assumption under certain physical conditions
(electron temperature, atomic density, etc.). Real galaxies could have a wide
range of nebular emission strength due to different physical states of the
gas, including geometry.

In this subsection, we take a more realistic approach to estimate the strength 
of nebular emission using the observed \lya\ line flux in our galaxy sample.
One advantage of our sample, other than the available spectroscopic redshifts, 
is the known \lya\ flux. Here we take advantage of this to estimate nebular 
emission, and incorporate the estimated nebular emission into our galaxy 
models. We first estimate the intrinsic \lya\ flux for each galaxy based on 
the observed \lya\ flux given in Paper I. \lya\ emission is complicated and 
largely reduced by the resonant scattering of \lya\ photons and the neutral 
IGM absorption. It is difficult to model \lya\ radiative transfer and predict 
the intrinsic \lya\ emission at high redshift. For a given apparent \lya\
luminosity, its intrinsic luminosity could have a range of values.
\citet{zheng10} showed that the distribution of the intrinsic to observed
flux ratio ($f_{\rm Ly\alpha}^{\rm int}/f_{\rm Ly\alpha}^{\rm obs}$) 
peaks at $\sim$4 in relatively bright galaxies at $z\sim5.7$. We thus assume 
$f_{\rm Ly\alpha}^{\rm int}/f_{\rm Ly\alpha}^{\rm obs}=4$ for our galaxies.
As we already mentioned, the {\tt GALEV} models do not include the \lya\ 
emission line. We link the \lya\ line flux to the model nebular line flux 
using \hb\ under Case B recombination, where the \lya\ to \hb\ ratio 
($f_{\rm Ly\alpha}^{\rm int}/f_{\rm H\beta}^{\rm int}$) is roughly 25 
($\rm Ly\alpha/H\alpha=8.7$ and $\rm H\alpha/H\beta=2.87$). 
The combination of the above two steps leads to an assumption that the ratio 
of the observed \lya\ to the intrinsic \hb\ flux ratio 
($f_{\rm Ly\alpha}^{\rm obs}/f_{\rm H\beta}^{\rm int}$) is 6.25, in the case 
of no dust extinction. 

In order to incorporate this ratio to the {\tt GALEV} models, we take each 
pair of the EM and NoEM model spectra in the whole parameter space from the 
previous subsection, and compute their nebular emission by subtracting the 
NoEM spectrum from the EM spectrum. We then scale the nebular emission 
(both line and continuum) to match the 
$f_{\rm Ly\alpha}^{\rm obs}/f_{\rm H\beta}^{\rm int}$ ratio (this ratio varies
with dust extinction, and it is taken into account in this step).
The details are as follows. We first take the \lya\ EW from Paper I. 
Based on the continuum of the model spectrum, the \lya\ EW, and the 
$f_{\rm Ly\alpha}^{\rm obs}/f_{\rm H\beta}^{\rm int}$ ratio, we calculate the 
\hb\ flux and EW for the model spectrum. Then the whole nebular emission 
spectrum is scaled and added to the model (continuum) spectrum. This model 
spectrum will have the required 
$f_{\rm Ly\alpha}^{\rm obs}/f_{\rm H\beta}^{\rm int}$ ratio 
during the SED fitting procedure. It does not guarantee that we will
obtain the observed \lya\ flux, because we only reserve its EW, the relative 
line flux to the continuum. Fortunately, there are no strong lines in the 
wavelength range covered by the non-IRAC data points (we did not use the bands 
that cover \lya). Therefore, the continuum at the wavelength of \lya\ is well 
modeled (see Figures 2 and 3). It ensures that the \lya\ flux in the
best-fitting model spectrum is very close to the observed flux. 
This has been tested for each object, and it is accurate to a few percent.

\begin{figure}
\epsscale{1.1}
\plotone{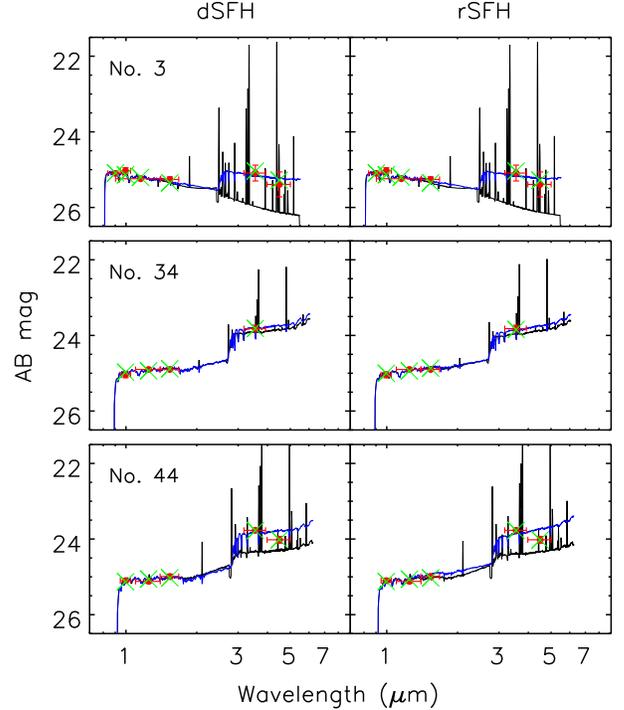}
\caption{Three examples of SED fitting with the \lya-EM models. The three
galaxies are the same as shown in Figure 2. The red points with error bars are
the observed photometric data points. The blue profiles are the NoEM
model spectra shown as a reference (same as those in Figure 2). The \lya-EM
model spectra are shown in black. The large green crosses represent
the photometric points predicted by the EM models.
The \lya-EM model spectra of No. 3 and
No. 44 show strong nebular emission lines as seen in Figure 2. The spectra of
No. 34 show much weaker lines compared to those in Figure 2, due to its weak
\lya\ emission line.}
\end{figure}

We next perform the SED modeling with the new EM models. We refer to these new 
EM models as `\lya-EM' models, and the original {\tt GALEV} EM models as
`GALEV-EM' models. Table 3 shows the results for the `\lya-EM' models with
dSFH and rSFH. Column 1 lists the galaxy sequence numbers. Columns 2--3 are 
stellar masses $M_{\ast}$ derived for the dSFH and the rSFH models, 
respectively. Columns 4--5 and Columns 6--7 are the derived ages and dust 
extinction for the two SFHs. Columns 8--9 give $\chi^2$ from the modeling.
Figure 3 illustrates the new SED modeling of the three galaxies shown in 
Figure 2. The blue profiles are the NoEM model spectra shown as a reference. 
The \lya-EM model spectra are shown in black. The \lya-EM model spectra for 
objects No. 3 and No. 44 still show strong nebular emission lines, as seen in 
Figure 2. The spectra for object No. 34, however, have much weaker lines 
compared to those in Figure 2. This is due to its weak \lya\ emission line 
(its rest-frame EW is only 8.6 \AA). Figure 4 compares the distributions of 
stellar mass, age, and $E(B-V)$ for the \lya-EM and NoEM models. These 
distributions are roughly consistent with those for the GALEV-EM models
shown in Figure 1. We will discuss this in detail in the next section.

\begin{figure}
\epsscale{1.1}
\plotone{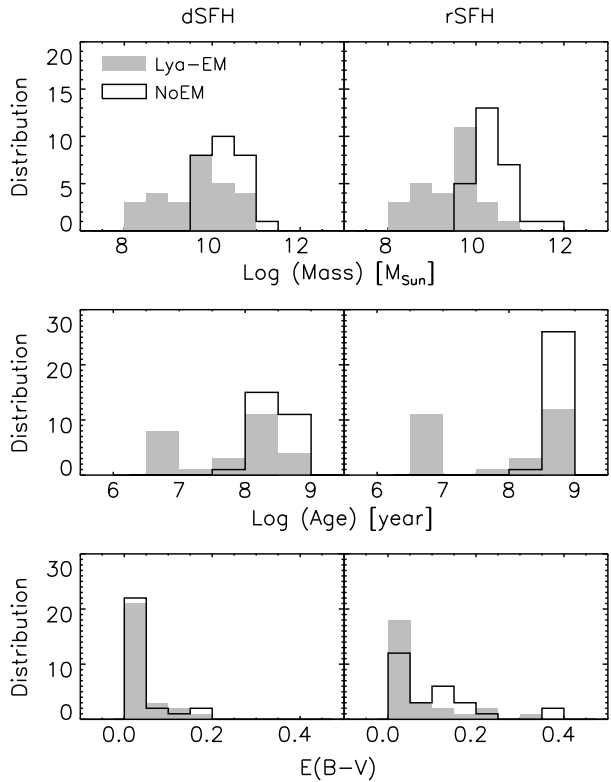}
\caption{Distributions of stellar mass, age, and $E(B-V)$ derived from the
\lya-EM models. Like Figure 1, the grey filled histograms show the results
from the \lya-EM models, and the black histograms show the results from the
NoEM models as references.}
\end{figure}

\section{RESULTS}

For galaxies with strong nebular emission, the results of SED fitting 
strongly depend on the relative strength of nebular emission. 
In the above section, we used
two methods to include nebular emission. One was to use the {\tt GALEV} models
with nebular emission (the GALEV-EM models), and the other one was to scale 
the nebular emission in the {\tt GALEV} models to match our observed \lya\ 
flux (the \lya-EM models). The latter one is a more realistic approach,
and consists of two assumptions: the ratio of the observed to intrinsic \lya\ 
flux $f_{\rm Ly\alpha}^{\rm int}/f_{\rm Ly\alpha}^{\rm obs}$ is 4, and 
the intrinsic \lya\ to \hb\ flux ratio 
$f_{\rm Ly\alpha}^{\rm int}/f_{\rm H\beta}^{\rm int}$ is 25 (Case B),
so that $f_{\rm Ly\alpha}^{\rm obs}/f_{\rm H\beta}^{\rm int}=6.25$. 
Each assumption involves non-negligible uncertainties. For the first 
assumption, the probability distribution function of 
$f_{\rm Ly\alpha}^{\rm int}/f_{\rm Ly\alpha}^{\rm obs}$ 
depends on the physical conditions of galaxies \citep{zheng10}. In addition,
it is apparently a function of redshift: it increases (observed \lya\ flux
decreases) towards higher redshifts as the neutral fraction of the IGM
increases. For the second assumption, real galaxies are more complex than the 
assumed ideal Case B recombination. 

Despite the uncertainties mentioned above, our assumptions are reasonable
(see more details in the discussion section).
As we will see, the \lya-EM and GALEV-EM models
produce roughly similar results on average, although there are large 
object-to-object variations due to the large range of the observed \lya\ EW. 
In order to explore the possible ranges of physical parameters, 
we vary the $f_{\rm Ly\alpha}^{\rm obs}/f_{\rm H\beta}^{\rm int}$ ratio by
a factor of two, i.e., we perform another two sets of SED modeling by 
assuming $f_{\rm Ly\alpha}^{\rm obs}/f_{\rm H\beta}^{\rm int}=3.125$
(referred to as EM-strong models), and
$f_{\rm Ly\alpha}^{\rm obs}/f_{\rm H\beta}^{\rm int}=12.5$ 
(referred to as EM-weak models). 
Based on the simulations of \citet{zheng10}, the majority of galaxies are
included in the range considered here. We will discuss nebular emission and
our assumptions in greater details in section 5.
%We discuss the details of the results in the next subsections.

\subsection{Stellar Mass}

\begin{figure}
\epsscale{1.1}
\plotone{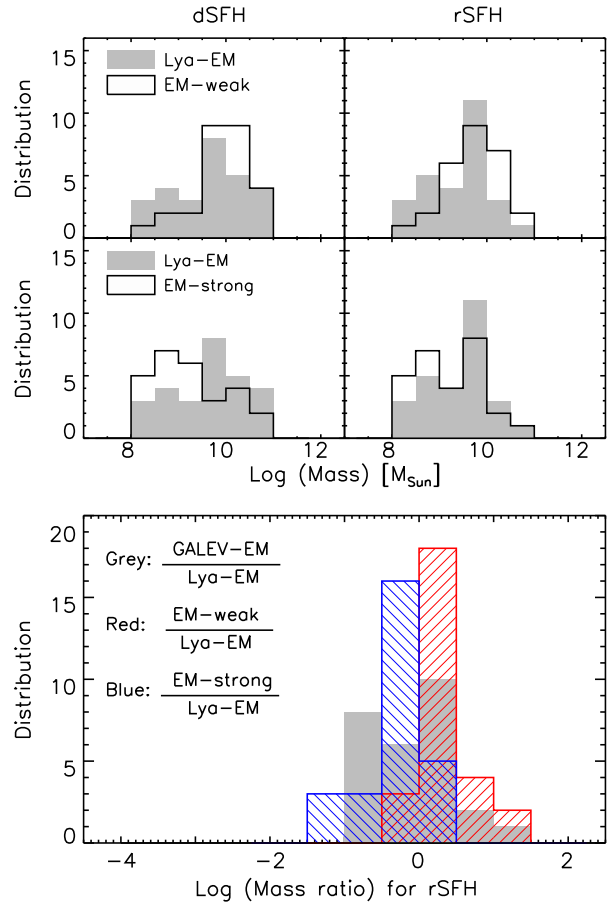}
\caption{Stellar masses derived from models with different nebular emission.
In the upper four panels, the filled grey histograms show the stellar mass
distributions from the \lya-EM models. The black unfilled histograms represent
the distributions from the EM-weak models and the EM-strong models,
which do not significantly deviate from those from the \lya-EM models.
The bottom panel shows the distributions of the mass ratios for rSFH.
The mass ratios of the GALEV-EM to \lya-EM models span a wide range due
to the wide distribution of the \lya\ EWs in our sample. The median value
is close to 1. Models with stronger (weaker) nebular emission
generally produce stellar populations with lower (higher) masses.
On the other hand, the results for most galaxies from three
EM models (\lya-EM, EM-weak, and EM-strong) are roughly consistent.
About 70--80\% galaxies in our sample have similar
stellar masses (within a factor of 3) from the three different EM models.}
\end{figure}

It is worth briefly discussing the bias of our sample before we discuss the
resulting stellar masses. The bias from sample selection was discussed in 
Paper I, and in Section 2 of this paper. The 42 galaxies in Table 1 are 
brighter than $\sim27.2$ AB mag in $J$. The 27 galaxies shown in Figures 1 and 
4 have detections in the IRAC 1 band (see also Table 1), so they are further 
limited by the 3.6 $\mu$m flux detection limit.
Because the 3.6 $\mu$m flux is closely related to stellar mass for $z\sim6$
galaxies, the sample of the 27 galaxies is biased towards the more massive
galaxies. This is the reason that the stellar masses derived from the NoEM 
models are mostly close to or higher than $10^{10}\,M_{\sun}$ (see the 
histograms in Figures 1 and 4).

\begin{figure}
\epsscale{1.1}
\plotone{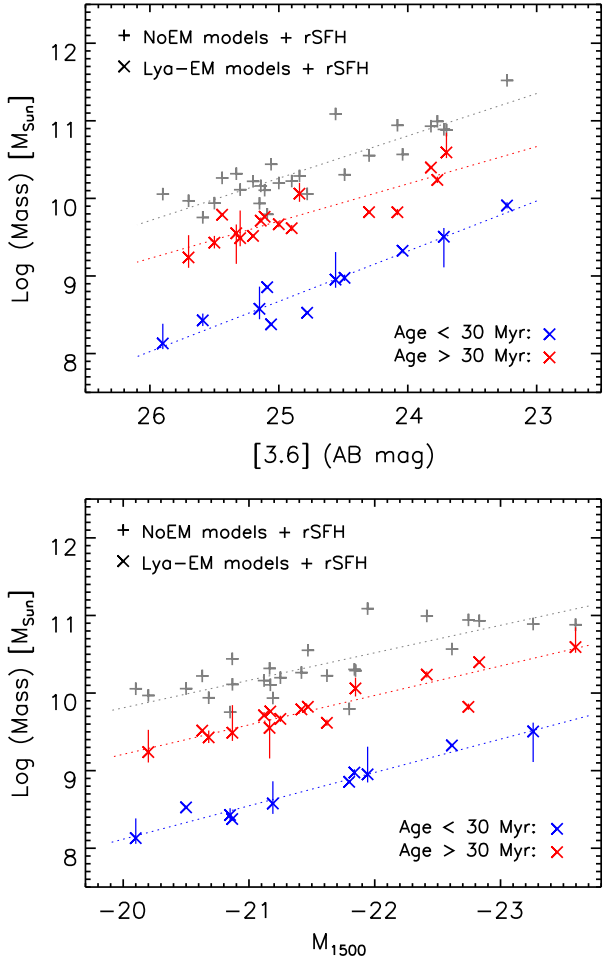}
\caption{Relations between stellar mass and brightness.
The grey pluses represent the galaxies in the NoEM models with rSFH. For the 
EM models with rSFH, the galaxies are divided into two subsamples, a `young' 
subsample (age $<$ 30 Myr; blue crosses) and an `old' subsample (age $>$ 30 
Myr; red crosses). The dotted lines are the best linear (in the log space) 
fits to the data points.
Upper panel: stellar mass as a function of apparent magnitude at 3.6 $\mu$m.
The masses derived from the NoEM models have a good correlation with the 3.6 
$\mu$m magnitude, because stellar mass is the amplitude of a model spectrum. 
For the EM models, the relation is also tight in either subsample, due to the 
reasons explained in section 4.1. But many galaxies are significantly below
the relation from the NoEM models, because of the presence of strong nebular 
emission in the IRAC 1 band. 
Lower panel: stellar mass as a function of absolute magnitude $M_{1500}$ 
at rest-frame 1500 \AA. For the 
NoEM models, there is a weak correlation in which stellar masses are higher 
in more UV luminous galaxies. This correlation is more obvious in the two 
subsamples for the EM models. The correlation may reflect the mass-SFR
relation seen at lower redshifts.}
\end{figure}

Stellar mass is thought to be the parameter that is least sensitive to model
assumptions, because it is directly measured from the scaling of galaxy model 
spectra \citep[e.g.][]{pap01,sha05}. When nebular emission is included, 
however, the measurement of stellar mass becomes less straightforward. 
Figure 5 compares the stellar masses derived from models with different 
nebular emission. The filled grey histograms show the stellar mass 
distributions from the \lya-EM models. The black unfilled histograms represent 
the mass distributions from the EM-weak models (top two panels) and the 
EM-strong models (middle two panels). The distributions of the 
EM-weak and EM-strong models do not significantly deviate from those 
for the \lya-EM models. The bottom panel shows the ratios of the stellar 
masses derived from different EM models. Although the mass ratios of the 
GALEV-EM to \lya-EM models span a wide range due to the wide distribution 
of the \lya\ EW in our sample, the median ratio is close to 1, suggesting 
that the assumptions we made for the \lya-EM models are reasonable. We use the 
EM-weak and EM-strong models to explore the possible mass ranges.
As expected, they show that models with stronger (weaker) nebular emission 
generally produce stellar populations with lower (higher) masses and younger 
(older) ages. On the other hand, the results for most galaxies from the three 
EM models (\lya-EM, EM-weak, and EM-strong) are roughly consistent.
The bottom panel shows that about 70--80\% galaxies in our sample have similar 
stellar masses (within a factor of 3) from the three different EM models.

The upper panel of Figure 6 shows the relation between the 3.6 $\mu$m flux and 
stellar mass derived from models with rSFH. There is a tight relation for 
the NoEM models (grey pluses), as explained above. For the EM models (the 
\lya-EM models here; crosses), the correlation still exists for the whole 
sample, but with a much larger scatter. The relation also significantly 
deviates from the relation for the NoEM models, due to the contamination of 
strong nebular emission in the IRAC 1 band. We divide these galaxies into two
subsamples: a `young' subsample (age $<$ 30 Myr; blue crosses) and an `old' 
subsample (age $>$ 30 Myr; red crosses) based on the age distributions shown 
in Figures 4 and 8. The old subsample consists of galaxies with ages of 
several hundred Myr, and the galaxies in the young subsample are usually 
younger than 30 Myr (see subsection 4.2). It is obvious that the relation 
between the 3.6 $\mu$m flux and stellar mass is tight in either subsample.
The linear Pearson correlation coefficients for the old and young subsamples
are --0.87 and --0.96, respectively. The upper panel of Figure 7 shows the
relation for dSFH, which is also obvious though less tight. The linear Pearson 
correlation coefficients for the two subsamples are --0.66 and --0.92, 
respectively. This relation is partially shaped 
by selection effects, i.e., these objects were selected in a relatively small 
parameter space. It may also be partially due to the following reasons.
For the old subsample, the 3.6 $\mu$m flux is still dominated by stellar 
emission, and thus reflects the stellar mass. In the young subsample, nebular
emission has a large contribution to the 3.6 $\mu$m flux. On the other hand,
these galaxies are within a small range of (very young) age, and their 
relative strength of the nebular to stellar emission does not span a wide
range (note that it is fixed for any GALEV-EM model SED). 
So the combination of the nebular and stellar emission still follow
the correlation, though the presence of nebular emission has largely reduced 
the required stellar masses. 

The lower panel in Figure 6 shows the relation between stellar mass and 
rest-frame UV luminosity $M_{1500}$, the absolute AB mag at 1500 \AA\ derived 
in Paper I). The $M_{1500}$ values have been corrected for dust extinction.
As for the upper panel, the pluses and crosses represent the NoEM and EM
models, and the red and blue crosses represent the old and young subsamples,
respectively. For the NoEM models, there is a weak correlation in which 
stellar masses are higher in more luminous galaxies. Such correlation is 
more tight in the two subsamples in the \lya-EM models. The linear Pearson 
correlation coefficients for the old and young subsamples are --0.92 and 
--0.99, respectively. The lower panel of Figure 7 shows the relation for dSFH, 
which is also obvious though less tight. The linear Pearson correlation 
coefficients for the two subsamples are --0.67 and --0.96, respectively.
This relation has been reported in previous studies of high-redshift galaxies 
with different SFHs, including constant SFH and slowly varying (rising or 
declining) SFHs \citep[e.g.][]{sta09,gon11,mcl11,cur13}. It is believed 
to be the high-redshift version of the mass-SFR relation (or the `so-called' 
main sequence of star-forming galaxies) found at lower redshifts, in which 
the SFR is higher in more massive galaxies, and the normalization of the 
relation at higher redshift is higher \citep[e.g.][]{dad07,elb07,noe07}.
Objects with extreme star-forming activity may significantly
deviate from the relation. In our sample the UV luminosity $M_{1500}$ reflects 
SFR. We will discuss the SFRs of our sample in section 4.4.
The tight relations seen in the figure could also be partially due
to the selection effects mentioned earlier.

From Figures 4, 5, and 6, all the four EM models show a wide range of stellar
masses in our galaxies, ranging from 
$M_{\ast}\sim10^8$ to $10^{11}$ $M_{\sun}$. In particular, a large fraction 
($\sim50$\%) of the galaxies have $M_{\ast}$ close to, or higher than 
$5\times 10^9\,M_{\sun}$, suggesting that very massive galaxies
already exist when the universe was only $\sim900$ Myr old. 

\begin{figure}
\epsscale{1.1}
\plotone{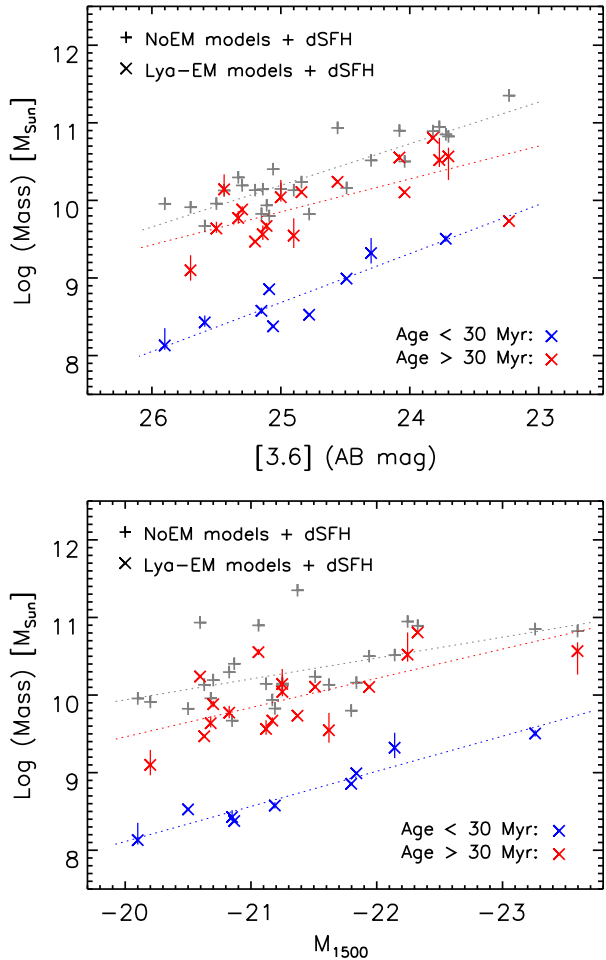}
\caption{The same as Figure 6, but for dSFH.}
\end{figure}

\subsection{Age}

\begin{figure} % f8
\epsscale{1.1}
\plotone{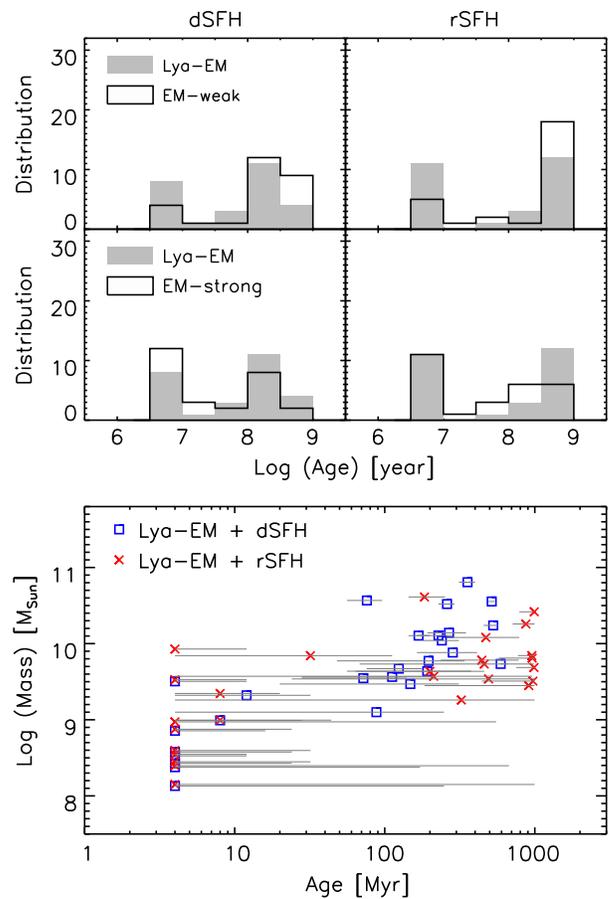}
\caption{Ages derived from models with different nebular emission. The filled
grey histograms show the age distributions from the \lya-EM models. The black
unfilled histograms represent the age distributions from the EM-weak
models (top two panels) and the EM-strong models (middle two panels).
The distributions of the different models look similar. The bottom panel shows
the relation between age and stellar mass derived from the \lya-EM models.
The grey horizontal bars indicate the $1\sigma$ uncertainties.
Note that we only show the uncertainty in the age direction and choose the 
best-fit mass; in reality there is a strong degeneracy between the two.
For the purpose of clarity, the symbols for the rSFH models have been shifted
by 0.02 along the y-axis.
Many galaxies are older than 200--300 Myr, with stellar masses close to, or
higher than $10^{10}\,M_{\sun}$. Meanwhile, a significant fraction of the
galaxies are dominated by extremely young stellar populations with ages of
several Myr.}
\end{figure}

As already mentioned, the age of a stellar population is usually poorly 
constrained from SED modeling, especially for high-redshift galaxies with only 
3--5 photometric data points. During our SED fitting, we allowed the age to 
vary between 4 and 1000 Myr.
Figures 1 and 4 show that the stellar populations derived from the NoEM 
models are mostly older than 100 Myr. The inclusion of nebular emission 
largely reduces the derived age, leading to extremely young (a few Myr) 
stellar populations in some galaxies. Figure 8 compares the ages derived from
models with different nebular emission. The filled grey histograms show the 
age distributions from the \lya-EM models. The black unfilled histograms
represent the age distributions from the EM-weak models (top two panels) 
and the EM-strong models (middle two panels). The distributions of the 
different models look similar. They all appear to show bimodal distributions. 
This bimodality could be real, but it could also be caused by selection 
effects (or sample bias) and modeling limitations. 
Our galaxies were selected to have strong \lya\ emission, which 
is biased towards younger populations. They were further limited by the IRAC 1
detections, which is biased towards higher stellar masses (older populations)
and/or those with strong nebular lines. A full exploration requires a
complete, mass-limited sample in this redshift range.
The other reason for the bimodality is modeling limitations. The minimum age 
allowed by our models is 4 Myr, so galaxies younger than 4 Myr would have a 
measured age of 4 Myr. In addition, these measured `young' ages usually have 
large uncertainties (typically 20$\sim$30 Myr), as shown in the bottom panel 
of Figure 8. Therefore, the actual age distribution could be more smooth.

Despite the selection effects and the measurement uncertainties of the ages,
we may divide our sample into two subsamples, based on Figures 4 and 8. One 
consists of old galaxies with ages of several hundred Myr, and the other one
includes young galaxies that are mostly only several Myr old. They are 
referred to as `old' and `young' subsamples in section 4.1. The bottom panel 
in Figure 8 shows the relation between age and stellar mass derived from the 
\lya-EM models. It suggests that slightly more than half of the galaxies are a 
few hundred Myr old. In particular, ages in some galaxies from the rSFH models
are older than 300--500 Myr. These galaxies usually have masses close to or 
higher than $10^{10}\,M_{\sun}$ (up to $10^{11}\,M_{\sun}$). 
These massive and old galaxies already appear to exist when the universe 
was only 0.8--1.0 Gyr old. 

Meanwhile, Figure 8 also shows the existence of extremely young galaxies in 
our sample. Both dSFH and rSFH models suggest that some galaxies are only 
several Myr old. These galaxies are usually less massive, with masses between 
$10^8$ and $3\times10^{9}\,M_{\sun}$. \citet{pir07} found that some LAEs at
$4<z<5.7$ in the Hubble Ultra Deep Field are very young (a few Myr old) with 
masses between $10^6$ and $10^8\,M_{\sun}$. The young galaxies in our sample 
are similar to these galaxies in terms of age, but are certainly more massive.
This implies that extremely young galaxies at high redshift already
have a wide range of stellar masses.

\subsection{Dust Extinction}

The bottom panels of Figures 1 and 4 display the distributions of dust
reddening $E(B-V)$. Unlike age and stellar mass, the $E(B-V)$ values derived 
from different models, including EM models and NoEM models, are consistent. 
They clearly suggest that the majority of the galaxies in our sample have 
little or no dust extinction. This is indeed expected. In Paper I, we 
reported that our galaxies have steep rest-frame UV slopes on average, with a 
median value of $\beta\sim-2.3$. Since the UV slope is very sensitive to dust 
extinction, such steep slopes already indicate little dust extinction. 
The results are also broadly compatible with recent observations
\citep[e.g.][]{wal12,ouchi13,ota14,wil15}. 
Several galaxies in our sample have
moderate dust extinction with $E(B-V)>0.1$. These galaxies are relatively
massive with masses higher than $10^{9}\,M_{\sun}$.

\subsection{Mass-SFR Relation}

\begin{figure} % f9
\epsscale{1.1}
\plotone{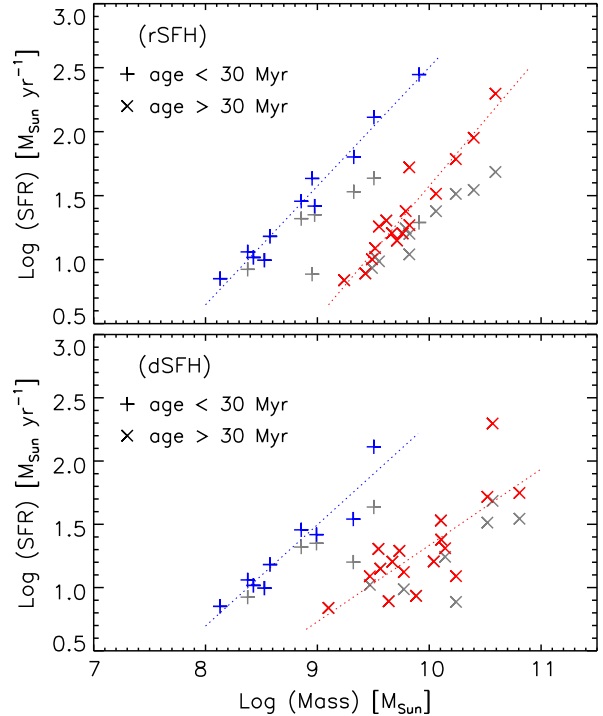}
\caption{The stellar mass - SFR relations, derived from the \lya-EM models 
with rSFH (upper panel) and dSFH (lower panel). The pluses and crosses 
represent the young and old subsamples, respectively. The grey symbols 
indicate the SFRs calculated from the UV continuum (without correction for 
dust extinction) in Paper I. The blue and red symbols indicate the SFRs
corrected for dust extinction. The dotted lines are the best linear (in 
the log space) fits to the data points.
The stellar masses are taken from the 
\lya-EM models with rSFH. The two subsamples show a tight mass-SFR relation, 
with slopes close to 1. The tight relation is likely caused by the 
combination of the intrinsic mass-SFR relation, selection effects, and
relatively small parameter space that two subsamples occupy.}
\end{figure}

In low-redshift star-forming galaxies, there is a correlation between stellar 
mass and SFR, in which SFR is higher in more massive galaxies
\citep[e.g.][]{dad07,elb07,noe07}. This relation is referred to as the 
`main sequence' of star-forming galaxies. It evolves with redshift so that
the normalization of the relation at higher redshift is higher (at least
for the redshift range $z\le 2-3$). The mass-SFR relation usually applies to 
massive galaxies with $M_{\ast} \ge 10^9$ $M_{\sun}$, corresponding to the old 
subsample in Figures 6 and 8. In addition, starburst galaxies such as
ULIRGs and SMGs may reside well above the relation. 

Figure 9 shows the mass-SFR relations for our galaxies, derived from the 
\lya-EM models with rSFH (upper panel) and dSFH (lower panel). The pluses and 
crosses represent the young and old subsamples, respectively. The grey symbols 
indicate the SFRs calculated from the UV continuum (without correction for 
dust extinction) in Paper I. We correct for dust extinction using the $E(B-V)$ 
values in Table 3 based on the Calzetti (2000) law. The results are shown as 
blue and red symbols. The correction is large only for the most massive 
galaxies in each subsample, which show moderate dust extinction as seen in 
section 4.3. The two subsamples both show a tight mass-SFR relation. 
In the upper panel for rSFH, the linear Pearson correlation coefficients for 
the old and young subsamples are 0.95 and 0.99, respectively. The coefficients
in the lower panel for dSFH are 0.77 and 0.95. 
The slopes are close to 1, consistent with the results from
simulations \citep[e.g.][]{finlator11}. It also agrees with the slopes found 
in lower-redshift galaxies \citep[e.g.][]{guo13,spa15}. The slope of $\sim1$
indicates that the specific SFRs (sSFRs) are similar within either subsample.
However, the average sSFR of the young subsample is much higher than 
that of the old subsample, so that the two subsamples are well separated in 
the mass-SFR diagram. This is because the two subsamples cover the similar 
range of SFRs, but the old subsample is ten times more massive on average.

The correlation in either subsample is tight, due to the combination of 
a few reasons, including the intrinsic mass-SFR relation, the selection 
effects, and the relatively small parameter space that either subsample
occupies. In particular, the correlation for the `young' subsample is very
tight, mainly because of their extremely young ages. The {\tt GALEV} model
outputs start from 4 Myr in steps of 4 Myr. The ages measured in the `young' 
subsample are mostly 4 Myr and 8 Myr, and thus these `young' galaxies have not 
evolved considerably. %, suggesting that the correlation seen for the `young' 
%subsample partially reflects the initial setups in the models.

Figure 10 shows the sSFR as a function of redshift, taken 
from \citet{mad14} (and references therein). It illustrates the efficiency of 
stellar mass growth in galaxies across cosmic time. The sSFR increases rapidly 
from the local Universe to $z\sim2$, and then flattens (or slightly climbs) 
towards higher redshifts. We focus on the high-redshift range. The $z\ge4$ 
data points in the figure were from two LBG samples of \citet{sta13} and 
\citet{gon14}. Our results are displayed as a blue circle (young subsample) 
and a red circle (old subsample). The horizontal error bars indicate the 
redshift range, and the vertical error bars indicate the $2\sigma$ range of 
the object number distribution (i.e., inclusion of 95\% of the objects). 
The average sSFR of the old subsample agrees well with the two previous studies 
\citet{sta13} and \citet{gon14}. Note that in these studies nebular
emission lines were also incorporated during their SED modeling.
In addition, the mass-SFR relation of the old subsample is well consistent 
with the simulation of \citet{finlator11}. All these suggest that the galaxies 
in our old subsample are `normal' star-forming galaxies at $z\ge6$. The sSFRs 
of the old subsample are roughly consistent with (or marginally higher than)
those at $z\sim2$, suggesting that the efficiency of stellar mass growth
did not change much in the most time of the first 3 Gyr.

\begin{figure}  % f10
\epsscale{1.1}
\plotone{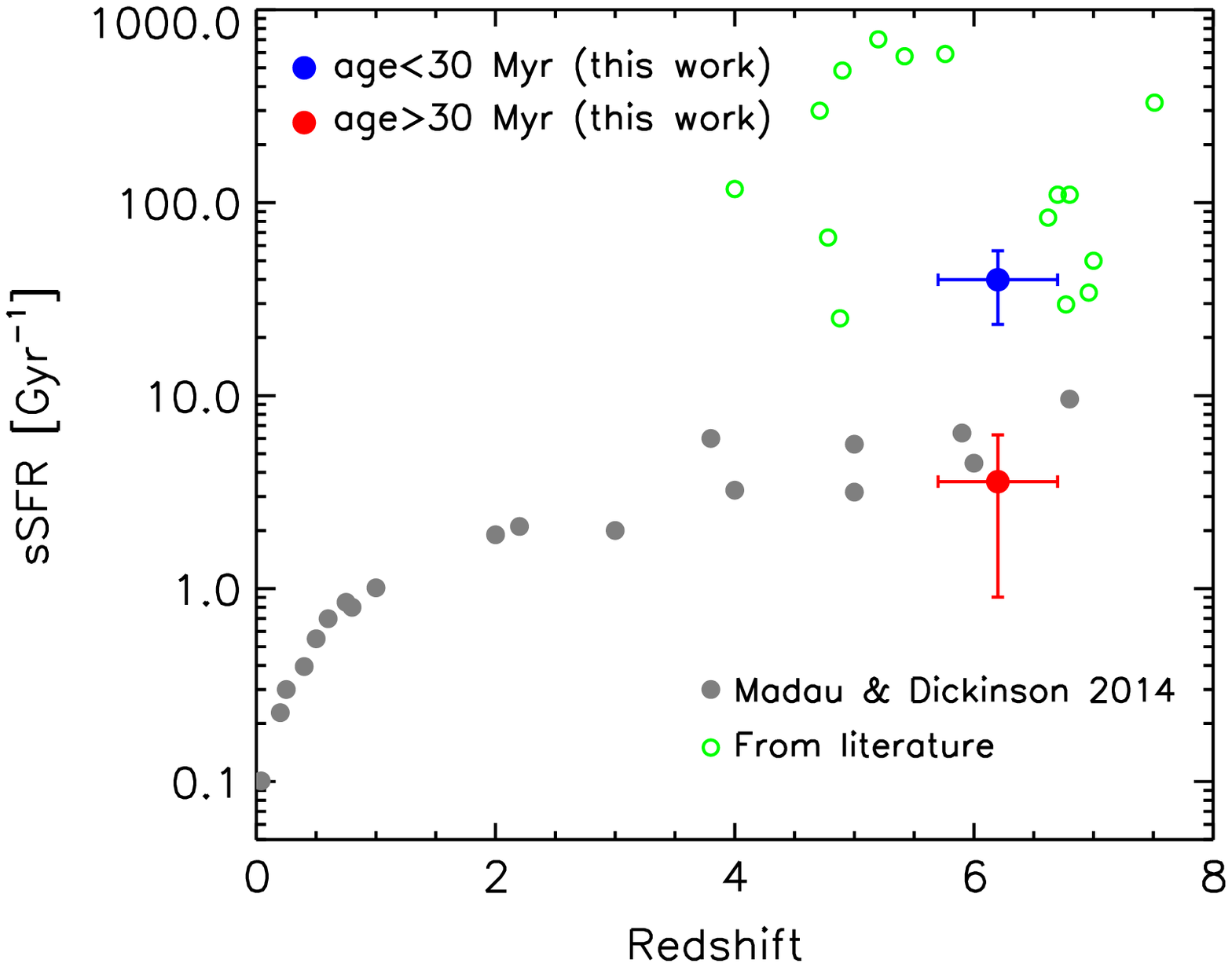}
\caption{The sSFR as a function of redshift. The grey circles represent the
data points taken from \citet{mad14} (and references therein). The open green
circles represent some of high-sSFR galaxies at $z>4$ taken from the
literature \citep{pir07,bow12,fin13,huang15}. Our results of 
the median sSFRs are displayed as a blue circle (young subsample) and a red 
point (old subsample). The horizontal error bars indicate the redshift range, 
and the vertical error bars indicate the $2\sigma$ range
of the object number distribution (i.e., inclusion of 95\% of the objects).
The sSFR of the old subsample is consistent with previous studies compiled by
\citet{mad14}, following
the main sequence of star-forming galaxies. The sSFR of the young subsample
is well above the main sequence, presumably due to starburst activity in
these galaxies. Such galaxies are not rare at high redshift, as seen from
the open circles.}
\end{figure}

The average sSFR of the young subsample is about ten times higher than the 
relation defined by previous studies. It has been clear that galaxies with 
strong starburst activity at $z\le2$ are well above ($\ge10$ times) the main 
sequence of star-forming galaxies \citep[e.g.][]{rod11,sar12}. So the young 
subsample seems to be the high-redshift counterparts of $z\le2$ starbursts. 
The fraction of such starbursts at $z\le2$ is very low, while in our sample 
this fraction is much higher, partly due to selection effects.
On the other hand, galaxies with very high sSFRs at high redshift are not
rare. Section 5.3 provides more discussion on this topic.

The bimodal distribution of ages and sSFRs seen in Figures 8 and 9 were 
largely explained by selection effects and modeling limitations in the above 
sections, though we cannot rule out the possibility of an intrinsic 
bimodality. Such bimodality has been reported for galaxies 
at $2.5<z<3.5$ by \citet{kaj10}, who found that the sSFRs in their low-sSFR 
and high-sSFR galaxies are 0.5--1.0 Gyr$^{-1}$ and $\sim$10 Gyr$^{-1}$, 
respectively. This suggests that the bimodality seen in our sample may 
partially reflect a real bimodal distribution of sSFRs.

\section{DISCUSSION}

\subsection{Testing \lya-EM Models}

It is clear that SED modeling of high-redshift galaxies can be largely 
affected by the presence of strong nebular emission. For galaxies at 
$6\le z<7$,
the existence of strong lines is often evidenced by their IRAC 1 flux excess
(compared to the IRAC 2 flux). This is because in this redshift range, the
IRAC 1 band covers some of the strongest lines such as \oiii, \ha, and \hb.
A significant fraction of galaxies at $z\ge6$, including photometrically
selected and spectroscopically confirmed galaxies, show a strong IRAC 1 flux
excess \citep[e.g.][]{gon10,mcl11,cur13}. In fact, many galaxies in our sample
also show such an excess. At $z>7$, galaxies with strong nebular lines
start to show red IRAC [3.6]--[4.5] colors \citep[e.g.][]{rob15,zit15}.

The contribution of nebular emission to the IRAC 1 and 2 bands can be large
or dominant. \citet{sta13} showed that the mean rest-frame EW of \ha\ at
high redshift is a few hundred \AA. \citet{smit15} showed some extreme cases
of strong emission lines by searching for $z=6.6-6.9$ galaxies with very blue
IRAC [3.6]--[4.5] colors. Over a small redshift range $z=6.6-6.9$, the IRAC
1 band covers \oiii\ and \hb, but the IRAC 2 band does not cover any
strong emission lines. The galaxies that they found have very large rest-frame
EW of \oiii+\hb\ in the range of 900 to $>2000$ \AA, with a median value of
$\sim1400$ \AA. In these galaxies line emission dominates the IRAC 1
photometry. Therefore, it is expected that the galaxies in our sample have
strong nebular lines.

Our analysis was mostly based on the \lya-EM models. When we computed the 
\lya-EM models from the GALEV-EM models in section 3, we scaled nebular 
emission lines using the observed \lya\ flux. Figure 11 shows the distribution 
of the derived scaling factors. They cover a range from
0.6 to 3.5, with a median value 1.6. It is expected that line emission derived
from the \lya-EM models is stronger than that from the GALEV-EM models,
because our galaxies were selected to have strong line emission.
The scaling factors span a relatively small range (within a factor of $\sim$2 
around the median value 1.6), suggesting that 
%our assumptions made for the \lya-EM models are reasonable, and 
the dynamic range of the EM-strong and EM-weak models is large 
enough to sample the majority of high-redshift galaxies. 

\begin{figure}  % f11
\epsscale{1.1}
\plotone{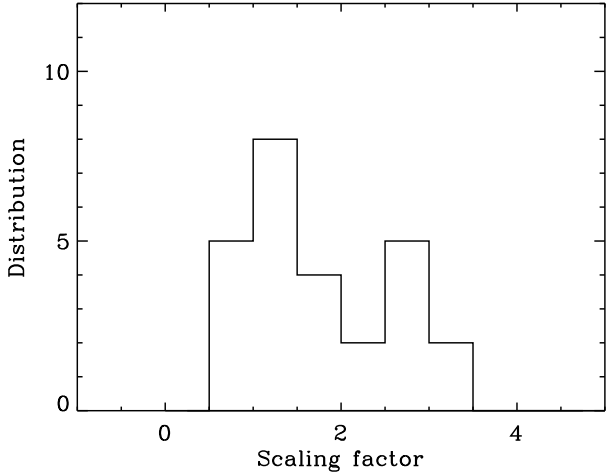}
\caption{Distribution of the scaling factors of nebular line strength derived
for the \lya-EM models. The \lya-EM models were computed from the GALEV-EM
models by scaling nebular emission lines to match the observed \lya\ flux
(section 3). The derived scaling factors cover a range from 0.6 to 3.5, with a
median value 1.6.}
\end{figure}

We may test our \lya-EM models using a few `special' galaxies (Figure 12) not 
in our sample. These galaxies are spectroscopically confirmed with 
measured \lya\ line flux, so that we can estimate nebular emission from their 
\lya\ emission based on the \lya-EM models. In addition, they are at certain 
redshift ranges so that one of the two IRAC bands covers strong nebular 
emission, but the other one does not. In this case, the difference of the 
photometry between the two bands roughly reflects the strength of nebular 
lines. The first two galaxies (a and b in Figure 12) are at $z=6.74$ and 
$z=6.76$ found by Cl{\'e}ment et al. (in preparation) and by \citet{huang15},
respectively. In these two galaxies, the IRAC 1 band covers \hb\ and \oiii, 
while the IRAC 2 band does not cover any strong lines. The other two galaxies
in Figure 12 are the one at $z=7.51$ from \citet{fin13} and the one at 
$z=7.73$ from \citet{oes15}. For these two galaxies, the IRAC 2 band covers 
\hb\ and \oiii, and the IRAC 1 band does not cover strong lines.

\begin{figure}  % f12
\epsscale{1.1}
\plotone{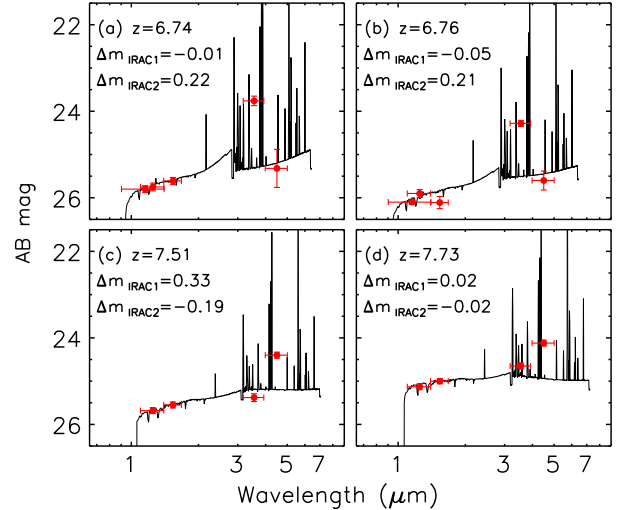}
\caption{SED modeling of four galaxies taken from the literature to test our
\lya-EM models. These galaxies are spectroscopically confirmed with measured
\lya\ line flux. They are at certain redshift ranges so that one of the two
IRAC bands covers strong nebular emission, but the other one does not.
We also show the difference between the observed and model photometry in the
IRAC bands ($\Delta m_{\rm IRAC1}$ and $\Delta m_{\rm IRAC2}$).
The fitting results for all the galaxies except (c) are good, and their
magnitude difference values are smaller than the corresponding photometric
uncertainties ($1\sigma$) in the IRAC 1 and IRAC 2 bands.
The best fit to galaxy (c) is not acceptable ($\chi_{r}^2\gg1$), due to
the reason that the nebular lines (scaled from \lya) in the model spectra are
not strong enough to account for the large difference between the IRAC 1 and
IRAC 2 photometry. This is because its \lya\ emission has been largely
attenuated by the highly neutral IGM at very high redshift.}
\end{figure}

We perform SED modeling for the four galaxies based on the \lya-EM models with 
rSFH. The fitting results for galaxies (a) and (d) are very good 
($\chi_{r}^2\le1$). 
We are not able to obtain acceptable results for galaxies (b) and (c), 
due to the reason that the nebular lines (scaled from \lya) in the model 
spectra are not strong enough to account for the large difference between
the IRAC 1 and IRAC 2 photometry. We note that the \lya\ emission line in
galaxy (b) is located on one of strong sky OH lines, and its flux measurement 
could be significantly affected, as pointed out by the discovery paper 
\citep{huang15}. Nevertheless, we re-model the SEDs for galaxies (b) and (c)
using the EM-strong models, and obtain a reasonable fit for galaxy (b). 
For galaxy (c), however, we fail to achieve an acceptable fit 
($\chi_{r}^2\gg1$). The best fitting results for the four galaxies are 
plotted in Figure 12, in which we also show the difference between the 
observed and model photometry in the IRAC bands ($\Delta m_{\rm IRAC1}$ and 
$\Delta m_{\rm IRAC2}$). For all the galaxies except (c), these magnitude 
difference values are smaller than the corresponding photometric uncertainties 
($1\sigma$) in the IRAC 1 and IRAC 2 bands.

Galaxy (c) has very weak \lya\ emission (compared to its other nebular lines),
as already noted by \citet{fin13}. At $z=7.51$, the IGM is much more neutral
than that at $z\sim6$, so \lya\ emission can be largely attenuated by the
neutral IGM or eaten by \lya\ damping wings \citep[e.g.][]{mir98}.
In other words, the ratio of intrinsic to observed \lya\ flux is a function of 
redshift, as we mentioned in the previous section. So our assumption about 
this ratio is no longer valid. 
Recent simulations suggest that the \lya\ damping wing owing to patchy 
reionization should be fairly uniform at a given redshift \citep{mes15}.
In this case, our ability to fit galaxy (d) but not (c) is not likely to 
reflect incomplete reionization. Instead, it may indicate scatter in the level 
of attenuation of self-shielded systems or in the intrinsic properties of the 
galaxies' interstellar media.

In summary, three out of the four galaxies can be well fit with our \lya-EM
or EM-strong models. We were not able to obtain a reasonable fit for the 
galaxy at $z=7.51$, due to the limitation (redshift coverage) of our models.
The tests above show that our models can provide reasonable SED modeling for
high-redshift galaxies with nebular emission taken into account.

\subsection{`Young' and `Old' Populations}

\begin{figure}  % f13
\epsscale{1.1}
\plotone{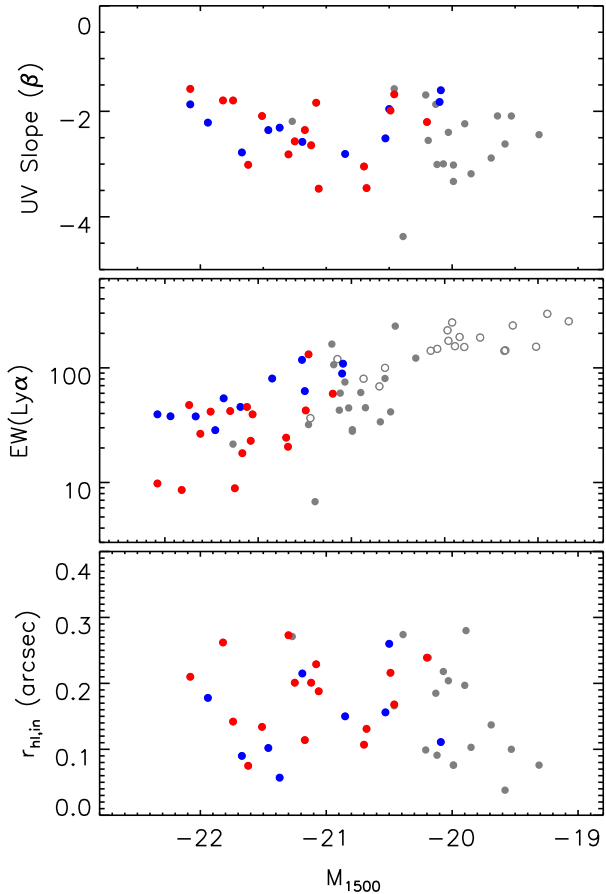}
\caption{Physical properties of the `young' and `old' subsamples as a function 
of rest-frame UV luminosity $M_{1500}$. The blue and red circles represent the 
`young' and `old' galaxies, respectively. The grey circles (including filled 
and open circles) represent the galaxies that are not included for analysis 
in this paper. The top panel shows the rest-frame UV slope $\beta$ derived
in Paper I. The two subsamples have similar mean slopes. The middle panel 
shows the rest-frame \lya\ EW derived in Paper I. The `young' galaxies have 
relatively higher \lya\ EWs than the `old' ones. The bottom panel shows 
the half-light radii $r_{hl,in}$ estimated in Paper II. The average radii of 
the two subsamples are roughly consistent.}
\end{figure}

In section 4 we identified two subsamples in our galaxies, one `old' subsample
with ages of several hundred Myr and one `young' subsample with ages of 
several Myr. Either subsample follows a tight stellar mass-SFR relation.
The mean sSFR of the `old' subsample is consistent with those in many 
high-redshift galaxies reported in the literature, while the mean sSFR of the 
`young' subsample is an order of magnitude higher. In Figure 13 we compare the 
two subsamples in the context of physical properties as a function of 
rest-frame UV luminosity $M_{1500}$. These physical quantities (including 
rest-frame UV slopes, \lya\ EWs, and half-light radii) were measured in Papers 
I and II. Their measurements are usually associated with considerable 
uncertainties (see Papers I and II), which are not plotted in the figure for 
the purpose of simplicity. The blue and red circles represent the `young' and 
`old' galaxies, respectively. The grey circles (including filled and open
circles) represent the galaxies that are not included for analysis in 
this paper.

The top panel of Figure 13 shows the rest-frame UV slope $\beta$ as a function
of $M_{1500}$. The median and standard deviation values of the slopes for the 
young and old subsamples are $-2.31\pm0.40$ and $-2.20\pm0.62$, respectively.
They are consistent. UV slopes 
are most sensitive to dust extinction. As we have seen in section 4.3, there 
is little dust extinction in these galaxies except for several of the most 
massive galaxies. So the `young' galaxies do not show bluer slopes than the 
`old' galaxies.

The middle panel of Figure 13 shows the rest-frame \lya\ EW as a function of 
$M_{1500}$. The median and standard deviation values of the EWs for the two 
subsamples are $54.2\pm30.7$ and $39.3\pm29.5$, respectively. The `young' 
galaxies have relatively higher \lya\ EWs than the `old' ones. A \lya\ EW 
is the ratio of the \lya\ line flux to the continuum flux. The \lya\ line
strength measures the strength of nebular lines, especially in our \lya-EM
models. We expect to see stronger nebular lines in younger systems.
So this panel reflects that a galaxy with stronger \lya\ line emission
tend to be younger.

The bottom panel shows the half-light radii $r_{hl,in}$ (at rest-frame 
$\sim$1800 \AA) as a function of $M_{1500}$. The radii have been corrected for 
PSF broadening with simulations. The median and standard deviation values of 
the radii for the two subsamples are $0.15\pm0.06$ and $0.20\pm0.06$, 
respectively. The old galaxies are marginally larger on average. However,
it is not straightforward to compare the sizes of the two subsamples. At lower 
redshift, there is a mass-size relation, in which more massive galaxies (the 
`old' subsample in our case) tend to be larger. On the other hand, the 
galaxies in our `young' subsample may have strong starburst activity
as indicated by their high sSFRs. A significant fraction of low-redshift
starbursts are mergers, which tend to have large sizes. 

\subsection{Comparison with Previous Studies}

There are a number of studies on the stellar populations of $z\ge6$ galaxies 
in the literature. The majority of these galaxies are photometrically selected 
LBGs. There was little study for $z\ge6$ LAEs. As we explained in Introduction,
almost all the known LAEs were discovered by ground-based telescopes, and did 
not have deep infrared observations. Our program provides the largest sample 
so far for studying stellar populations in spectroscopically confirmed LAEs at 
$z\ge6$. 

Direct comparison with previous studies is very difficult if not impossible,
because different studies use different galaxy samples (selected from 
different datasets with different selection criteria) and SED modeling 
methods. We mainly focus on two questions: whether extremely young populations
have been reported, and whether previously
reported sSFRs cover a wide range that agrees with our `young' and `old'
subsamples. As we emphasized earlier, age is poorly constrained from SED
modeling, so here we broadly define `extremely young galaxies' as those
with the best-fitting ages younger than $\sim30$ Myr, like the galaxies in our 
`young' subsample.

In the previous studies of photometrically-selected LBGs, extremely
young galaxies have been rarely reported. The majority of these LBGs tend to
have relatively old and mature populations with ages of a few hundred or 
several tens Myr, from SED fitting without nebular emission taken into account. 
When nebular emission were considered (especially in models with 
smoothly-varying SFHs), however, extremely young populations were required
to explain the SEDs of some LBGs \citep{sch10,mcl11,cur12,huang15}. 
For example, \citet{mcl11} selected a sample
of $z>6$ LBGs from $HST$ deep fields, and found that these galaxies were 
mostly a few hundred Myr old if nebular emission was not included in their SED 
modeling. When nebular emission was added, about 25\% of their LBGs with IRAC 
1 detections were found to be extremely young, with a median age of 9 Myr.

In the previous studies of high-redshift LAEs, extremely young galaxies were
found to be common. For example, \citet{ono10} stacked a large number of
photometrically-selected LAEs at $z\sim5.7$ and 6.5, and performed SED
modeling on the stacked LAEs. Despite that the measurements from stacked data 
could be unreliable \citep[e.g.][]{var14}, they found that these LAEs have 
ages of only 1-3 Myr, little dust extinction, and strong nebular emission.
In the sample of \citet{pir07}, there are nine spectroscopically confirmed 
LAEs at $z\sim5$. Most of them were found to be younger than 10 Myr.
These results are quite consistent with ours.

The sSFRs in our galaxies span a quite large range. Their values are about
3--4 Gyr$^{-1}$ in the `old' subsample, which is consistent with previous
studies such as \citet{cur12}, \citet{sta13}, and \citet{gon14}. 
The sSFRs in our `young' subsample are about an order of magnitude higher,
likely due to starburst activity in these galaxies. Such high sSFRs are indeed
also common in previous studies. For example, in a sample of seven secure 
high-redshift LBGs by \citet{bow12}, three of them were found to have sSFRs
around 4--5 Gyr$^{-1}$, and another three have sSFRs close to or higher than 
30 Gyr$^{-1}$. These numbers roughly agree with those found in our `old' and
`young' subsamples. Some studies have reported even stronger starbursts
in $z>6$ galaxies, with sSFRs between one hundred and several hundred 
Gyr$^{-1}$ \citep[e.g.][]{ono10,fin13,huang15}. In Figure 10, the green 
circles represent some of high-sSFR galaxies from \citet{pir07}, 
\citet{bow12},
\citet{fin13}, and \citet{huang15}. Galaxies ($4<z<6$) from \citet{pir07} all 
have very high sSFRs, and are included in Figure 10. All these above suggest 
that strong starburst activity is common in very high-redshift galaxies.

\subsection{LAEs and LBGs}

\begin{figure} % f14
\epsscale{1.1}
\plotone{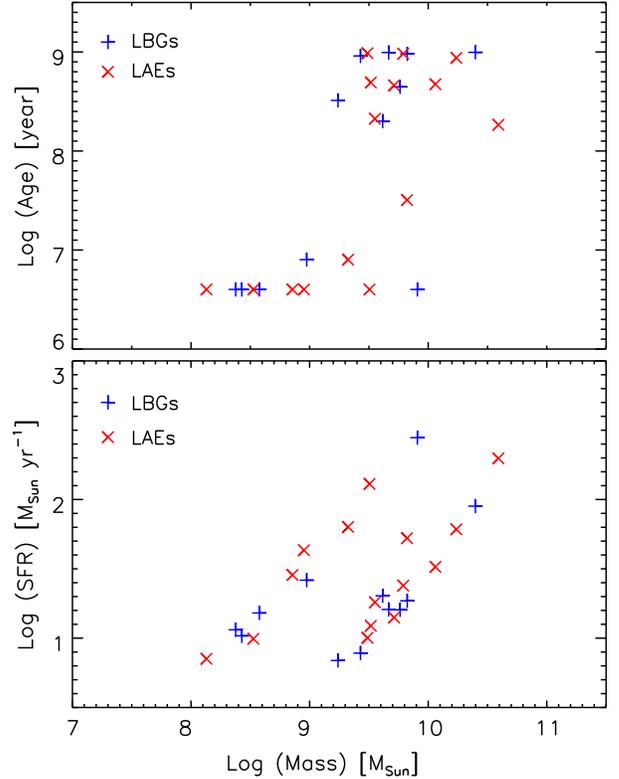}
\caption{The measured ages and SFRs as a function of stellar mass for the LAEs
and LBGs in our sample. The minimum age allowed in our models is 4 Myr, and
some ages in the upper panel are at this limit. The distributions of the LBGs 
and LAEs in the age-mass and SFR-mass diagrams are indistinguishable. 
The number ratios of the LAEs to LBGs 
agree with each other in the `young' and `old' subsamples.}
\end{figure}

In this series of papers (including this paper and Papers I and II), LAEs are 
defined as galaxies found by the narrow-band (or \lya) technique, and LBGs 
are defined as galaxies found by the dropout technique. As we already pointed 
out in Papers I and II, this widely-used classification only reflects the
methodology that we apply to select galaxies. It does not mean that the two
types of galaxies are intrinsically different. Another definition of LAEs 
is based on the \lya\ EW, e.g., a galaxy is a LAE if its \lya\ EW is greater 
than 20 \AA. This definition is physically more meaningful, but 
observationally difficult, because one can easily obtain a flux-limited 
sample, not a EW-limited sample. In addition, it is meaningless for
broad-band selected galaxies. So we use the former definition.

It is not entirely clear whether high-redshift LAEs and LBGs represent 
physically different populations. Direct comparison between LAEs and LBGs is 
difficult because of the very different target selection procedures.
As we already explained in Papers I and II, the LAEs in our
original sample of 67 galaxies are composed of a well-defined sample in terms 
of \lya\ flux. On the other hand, the LBGs in the sample only represent the 
LBGs with strong \lya\ emission since they are spectroscopically confirmed.
In Papers I and II, we compared the LAEs and LBGs in our sample in great
details, and found that the two populations are indistinguishable in all 
aspects of physical properties that we considered, including the \lya\ 
emission strength, UV continuum properties, sizes, and morphology, etc.

Figure 14 shows the distributions of the measured stellar masses, ages, and
SFRs for the LAEs and LBGs in our sample of 27 galaxies used in this paper.
The blue pluses and the red crosses represent the LBGs and LAEs, respectively.
%The upper panel shows the relation between age and stellar mass, and 
%the lower panel shows the relation between SFR and stellar mass.
The distributions of the 
LBGs and LAEs in the age-mass and SFR-mass diagrams are indistinguishable. 
In particular, there are 6 LAEs and 5 LBGs in the `young' subsample, and
there are 9 LAEs and 7 LBGs in the `old' subsample. The number ratios of 
the LAEs to LBGs nicely agree with each other in the two subsamples.
All these are consistent with our previous conclusion that the LAEs and LBGs 
in our sample have common properties, suggesting that LAEs are a subset of 
LBGs with strong \lya\ emission lines. The conclusion is also consistent with
recent simulations \citep[e.g.][]{day12,gar15}.

\subsection{Prospect for Spectroscopic Follow-up with {\em JWST}}

Currently there are two ways to assess the strength of rest-frame
optical nebular emission lines and use that information to derive the
physical properties of underlying stellar population in $z>6$
galaxies: (1) to observe galaxies in specific redshift ranges where
one of the IRAC band is line-free (e.g., $6.6<z<6.9$ galaxies as
studied by Smit et al. (2014)), and (2) to model emission-line
strengths based on the measured Ly$\alpha$ lines (this work).  The
first method is more direct and accurate, but because of the stringent
constraints on redshift, the number of galaxies that can be studied is
rather small.  The second method suffers from relatively large uncertainties, 
but can be applied to a much larger sample of $z>6$ galaxies as we have
shown in this paper.

Although the above is the best we can do at the moment, we note that we
are actually on the verge of a breakthrough as
the {\em JWST} becomes available and changes the emphasis from
photometry to spectroscopy. Figure 15 shows the flux distributions of
the [\ion{O}{3}] 5007 \AA\ and H$\alpha$ lines predicted by the
modeling described in this paper.  The figure shows that we should be
able to detect almost all these lines with a line-flux sensitivity of
$\sim10^{-17}$ ergs cm$^{-2}$ s$^{-1}$.  Based on the {\em JWST}
Prototype Exposure Time Calculator (ETC), such a sensitivity will be easily
reachable with NIRSpec.  For example, to achieve a 5$\sigma$ detection
(per resolution element) of an [\ion{O}{3}] 5007\AA\ line at $z=6.5$
with a line flux of $10^{-17}$ ergs cm$^{-2}$ s$^{-1}$, the required
integration time will be only $\sim500$ seconds.
The corresponding integration time for H$\alpha$
(i.e., at $z=6.5$ and with a line flux of $10^{-17}$ ergs cm$^{-2}$
s$^{-1}$) will be $\sim800$ seconds.
These numbers clearly indicate that with {\em JWST}/NIRSpec, it will
be easy to detect [\ion{O}{3}] and H$\alpha$ lines with the
brightest $z>6$ galaxies like those we have studied here.  In fact,
NIRSpec has the sensitivity to detect much fainter lines, making these
bright $z>6$ SDF galaxies ideal targets for a detailed NIRSpec
spectroscopic study.  In several years, such a study will be able to
test the validity of the results presented here, and will undoubtedly
make many more interesting discoveries.

\begin{figure}  % f15
\epsscale{1.1}
\plotone{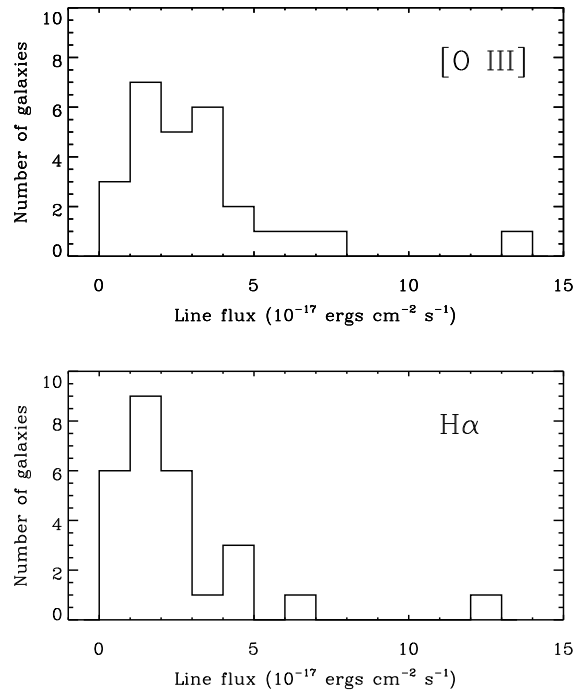}
\caption{Flux distributions of the [\ion{O}{3}] 5007 \AA\ and H$\alpha$ lines 
predicted by the modeling described in this paper. The flux was calculated
based on the {\em JWST} prototype ETC (version P1.6), available at 
http://jwstetc.stsci.edu. The following parameters were used with the 
NIRSpec ETC: Filter/Grating $=$ G395H/F290LP; $R=2700$; MSA shutter $=$ 
0\farcs2 $\times$ 0\farcs45; Flat continuum in F$\nu$; FWHM = 40 \AA\ $\sim$ 
300 km s$^{-1}$ at 3.755 $\mu$m for [\ion{O}{3}], or FWHM = 50 \AA\ $\sim$ 
300 km s$^{-1}$ at 4.922 $\mu$m for H$\alpha$. 
The continuum normalization was set to be 30 AB 
mag at 3.6 $\mu$m, so that the S/N calculation is essentially driven by the
line flux. With {\em JWST}/NIRSpec, it will be easy to detect [\ion{O}{3}] and 
H$\alpha$ lines in bright $z>6$ galaxies.}
\end{figure}

\section{SUMMARY}

This paper is the third in a series presenting the physical properties of a
large sample of 67 spectroscopically confirmed galaxies at $z\ge6$.
The sample consists of 51 LAEs at $z\simeq5.7$, 6.5, and 7.0, and 16 LBGs at
$5.9\le z\le6.5$. They have deep optical imaging data from Subaru, near-IR
data from $HST$, and mid-IR data from $Spitzer$. In this paper, we have
reported a detailed study of stellar populations in these galaxies.
We have focused on a subsample of 27 galaxies with $Spitzer$ IRAC 1 
detections at 3.6 $\mu$m. This subsample represents luminous and massive 
galaxies with strong \lya\ emission at $z\ge6$. We used the wealth of the 
multi-band data and the secure \lya\ redshifts and flux to model the 
SEDs of the 27 galaxies and characterize their stellar populations.

We used the {\tt GALEV} evolutionary synthesis models with nebular continuum 
and line emission to mainly constrain three physical parameters: age, stellar 
mass, and dust extinction. We adopted two representative SFHs, 
an exponentially declining SFH (dSFH) and a smoothly rising SFH (rSFH). 
We mostly used the latter two SFHs in our analysis. 
In order to incorporate nebular emission, we scaled the nebular emission from
the {\tt GALEV} models to match the observed \lya\ flux (the \lya-EM models)
under two simple assumptions. With the \lya-EM models, we were able to nicely
break the strong degeneracy of model spectra between young galaxies with 
prominent nebular emission and mature galaxies with strong Balmer breaks.

Our best-fitting results show that the galaxies in our sample has a wide range 
of SED ages from several Myr to a few hundred Myr. They also have a wide range 
of stellar masses from $\sim10^8$ to $\sim10^{11}\,M_{\sun}$. Interestingly, 
the distribution of the measured ages (despite the large uncertainties) appear 
to be bimodal, likely due to selection effects and modeling limitations 
(though we cannot rule out the possibility of an intrinsic bimodality). 
Based on this bimodality, we divided the galaxies into two subsamples: an `old'
subsample and a `young' subsample. The `old' subsample mainly consists of 
galaxies older than 100 Myr, with stellar masses higher than $10^9\,M_{\sun}$. 
Many galaxies are older than 300--500 Myr. The `young' subsample consists of 
galaxies younger than 30 Myr (usually several Myr old). These galaxies are 
less massive, with masses ranging between $\sim10^8$ and 
$\sim3\times10^9\,M_{\sun}$. 
The majority of the galaxies both subsamples show little or no dust 
extinction, as already hinted by their steep rest-frame UV slopes.

Both subsamples show a correlation between stellar mass and SFR, but with 
very different normalizations. The mean sSFR of the `old' subsample is 
about 3--4 Gyr$^{-1}$, consistent with the mass-SFR relation defined by 
previous studies. The mean sSFR of the `young' subsample is an order of 
magnitude higher. Such higher sSFRs have also been frequently reported in 
previous studies. They are likely due to starburst activity in these galaxies. 
Finally, the LAEs and LBGs in our sample are indistinguishable in all 
physical properties that we have considered, suggesting that 
LAEs are a subset of LBGs with strong \lya\ emission lines.

\acknowledgments

We acknowledge the support from a 985 project at Peking University.
L.J., S.C. and E.E. also acknowledge the support from NASA through awards 
issued by STScI ($HST$ PID: 11149,12329,12616) and by JPL/Caltech ($Spitzer$ 
PID: 40026,70094). We would like to thank R. Ryan for his advice on IRAC 
photometry and using {\tt iGALFIT}. We also thank J. Rhoads and Z. Zheng
for their insightful comments, and thank P. Madau and P.A. Oesch for 
providing us the data used for Figure 10 and Figure 12.

{\it Facilities:}
\facility{$HST$ (NICMOS,WFC3)},
\facility{$Spitzer$ (IRAC)},
\facility{$Subaru$ (Suprime-Cam)}

\end{document}